\documentclass[lettersize,journal]{IEEEtran}
\ifCLASSINFOpdf
\else
\fi
\usepackage{stfloats}
\usepackage{cite,amssymb}
\usepackage{mathrsfs}
\usepackage[pdftex]{graphicx}
\usepackage{algorithm}
\usepackage[cmex10]{amsmath}
\usepackage{textcomp}
\usepackage{algpseudocode}
\usepackage{array,balance,bm}
\usepackage{amsmath}
\usepackage{nicematrix}
\usepackage{arydshln} 
\usepackage{lipsum}

\ifCLASSOPTIONcompsoc
\else
\fi
\usepackage{url}
\usepackage{amsfonts}
\usepackage{exscale}
\usepackage{relsize}
\usepackage{multicol,color}
\usepackage{array}
\usepackage{arydshln}
\usepackage{amsmath}

\newcommand{\be}{\begin{equation}}
	\newcommand{\ee}{\end{equation}}

\allowdisplaybreaks[4]

\begin{document}
\bstctlcite{setting}
	\title{RIS-Aided Integrated Sensing and Communication Systems under Dual-polarized Channels}
	
\author{\IEEEauthorblockN{Dongnan Xia, Cunhua Pan, \IEEEmembership{Senior Member, IEEE}, Hong Ren, \IEEEmembership{Member, IEEE}, Zhiyuan Yu, Yasheng Jin and Jiangzhou Wang, \IEEEmembership{Fellow, IEEE}}\\
	
	\thanks{D. Xia, C. Pan, H. Ren, Z. Yu, Y. Jin and J. Wang are with National Mobile Communications Research Laboratory, Southeast University, Nanjing, China. (e-mail:\{dnxia, hren, cpan, zyyu, yashengjin, j.z.wang\}@seu.edu.cn).
	}  
	\thanks{Corresponding author: Cunhua Pan and Hong Ren. }
}

\maketitle
\begin{abstract}
	This paper considers reconfigurable intelligent surface (RIS)-aided integrated sensing and communication (ISAC) systems under dual-polarized (DP) channels. 
	Unlike the existing ISAC systems, which ignored polarization of electromagnetic waves, this study adopts DP base station (BS) and DP RIS to serve users with a pair of DP antennas.
	The achievable sum rate is maximized through jointly optimizing the beamforming matrix at the DP BS, and the reflecting coefficients at the DP RIS.
	To address this problem, we first utilize the weighted minimum mean-square error (WMMSE) method to transform the objective function into a more tractable form, and then an alternating optimization (AO) method is employed to decouple the original problem into two subproblems. 
	Due to the constant modulus constraint, the DP RIS reflection matrix optimization problem is addressed by the majorization-minimization (MM) method.
	For the DP beamforming matrix, we propose a penalty-based algorithm that can obtain a low-complexity closed-form solution. 
	Simulation results validate the advantage of deploying DP transmit array and DP RIS in the considered ISAC systems.
\end{abstract}


	\section{Introduction}
	
	\IEEEPARstart{T}{he}  growing needs of wireless communication networks have increased the requirement to share spectrum between radar and communication systems\cite{ma2020joint,lu2024integrated}.
	Spectrum congestion is a significant issue that can reduce the capacity of nearby wireless communication bands.
	Integrated sensing and communication (ISAC), which allows for the sharing of spectrum between wireless communications and radar systems, is regarded as a key technology for 6G\cite{zheng2019radar,liu2018mimo}.

	The design of appropriate waveforms is crucial for achieving an efficient ISAC system.  Research in this area can be generally categorized into three approaches\cite{zhou2022integrated}: communication-centric waveform design (CCWD), sensing-centric waveform design (SCWD), and joint waveform optimization and design (JWOD). 
	CCWD modifies traditional communication waveforms, such as orthogonal frequency division multiplexing (OFDM) \cite{zhu2009chunk,zhu2011chunk} and orthogonal time frequency space (OTFS), to include sensing capabilities \cite{wu2021otfs,shi2022device}. However, this approach can weaken sensing performance because communication signals tend to vary frequently.
	SCWD methods focus on modifying sensing waveforms to carry communication signals\cite{hassanien2015dual}. While this approach offers implementation simplicity, it often results in limited data transmission rates and is primarily suitable for specific application scenarios. 
	 In contrast,  JWOD develops new waveforms for specific scenarios\cite{zhong2023joint,liu2021dual}, providing a new solution that achieves dynamic trade off between communication and sensing functions, thereby potentially enhancing the system's effectiveness.

	 The performance of the existing ISAC systems is often constrained by various limitations, particularly bandwidth restrictions. To enhance system performance, researchers have  explored additional physical characteristics of the channel.  Exploiting the polarization domain presents an effective solution to increase the number of antennas at the base station (BS) within a confined space which effectively doubles the number of antennas without increasing the array's physical dimensions\cite{nabar2002performance}.
	 The use of dual-polarized (DP) antennas, which leverage two orthogonal polarizations to establish parallel transmission paths, enables a two-fold increase in terms of capacity compared to single-polarized (SP) systems\cite{coldrey2008modeling}.

	 The adoption of DP antennas improves the spatial efficiency of antenna array but also increases system complexity. 
	 The channel modeling for DP channels is more complicated than conventional SP channels. 
	 Geometry-based channel models for DP small-scale MIMO systems have been developed through measurement campaigns\cite{shafi2006polarized}. In contrast, analytical models for single-user DP MIMO systems have been derived from comprehensive experimental surveys \cite{oestges2008dual}. The diversity of channel models highlights that polarization-related properties vary significantly across scenarios, precluding a one-size-fits-all approach.

	 Several depolarization effects, including cross-polar discrimination (XPD), cross-polar correlation (XPC), and cross-polar isolation (XPI), play a critical role in signal propagation within DP channel models \cite{ozdogan2022massive}. XPD quantifies a channel's ability to maintain polarization integrity between orthogonally polarized waves, typically showing higher values in line-of-sight (LoS) outdoor environments and lower values in dense scattering conditions \cite{Oestges2008}. XPC, which characterizes the mutual interference between orthogonal polarization components and is often represented by a 2×2 matrix \cite{coldrey2008modeling}, also affects system performance. XPI exists as a hardware-related parameter measuring polarization leakage at the antenna, and is typically negligible compared to environmental depolarization effects, with values generally exceeding 30 dB, and can be calibrated in digital baseband processing \cite[Ch. 8]{stutzman2018polarization}.

	 	 DP MIMO systems have been extensively studied, from basic 2x2 setups \cite{nabar2002performance} to   commercial MIMO configurations \cite{de2019massive,swaminathan2012performance,park2015multi}.
	 Specifically, \cite{de2019massive}  studied DP massive MIMO-NOMA systems, offering two precoder designs and demonstrating performance improvements over SP systems. \cite{swaminathan2012performance} proposed a  singular value decomposition (SVD) precoding method for dual-polarization MIMO systems, achieving better performance than traditional methods.

	 In addition to communications, polarization technology plays a crucial role in radar systems. 
	  The polarization effect provides rich target information through a specific mapping transformation that depends on target characteristics such as posture, size, structure, and material properties \cite{sinclair1950transmission,kennaugh1952polarization}. Research in this field encompasses several key areas: precise polarization information acquisition, polarization-sensitive array processing, target scattering characteristics modeling \cite{ergul2008efficient}, anti-interference technology \cite{moisseev2000doppler}, and target classification and recognition. As highlighted in \cite{chen2014modeling}, these polarization mechanisms in synthetic aperture radar demonstrate significant potential in enhancing radar detection, anti-interference, and overall sensing capabilities.

	Although both sensing and communication functions may benefit from the DP techniques, they still
	suffers from challenging propagation environments that include signal blockages, which  deteriorate target sensing performance. A promising solution to this issue is the use of reconfigurable intelligent surfaces (RIS), which can manipulate the wireless propagation environment efficiently with minimal power consumption and hardware costs \cite{pan2022overview,zhi2022ris,zhang2024secure,liu2024joint,wan2024reconfigurable}. 	Initially, RIS applications focused on improving communication capabilities, while direct links between transceivers and targets are used for sensing \cite{zhu2022resource,wang2021joint,song2024overview}. 
	In this setup, the RIS helps enhance the signal strength for UEs, creating a robust BS-RIS-users link. 
	Additionally, RIS can also be used to enhance sensing ability \cite{jin2024reconfigurable,yu2023active}.
	
	  In addition to polarization at the antennas, DP RIS was also considered in \cite{munawar2023dual,ramezani2023dual,han2021dual,mohamed2022energy,zheng2024ris}. For instance, \cite{mohamed2022energy} introduced a method to improve downlink transmissions by optimizing beamforming techniques, enhancing energy efficiency using a DP aerial surface.  \cite{zheng2024ris}  investigated how large the  DP RIS can be used  to ensure  better performance for DP MIMO.

	 However, despite the extensive contributions of polarization in communication and sensing, to the best of our knowledge, the potential of polarization in ISAC has not been fully explored.
	 In our study, we  consider a BS equipped with DP antennas that can emit a pair of orthogonally polarized waves to perform communication and sensing tasks simultaneously. Accordingly, users are equipped with co-located DP antennas instead of traditional SP antennas. 
	 We use different polarization waves to detect different directions at the same time. In this way, we can analyze the scattering characteristics of the target in response to single polarization waves.
	 Therefore, introducing polarization into ISAC can not only improve communication efficiency  but also significantly improve sensing capabilities. Moreover, the introduction of the DP RIS technology can further improve communication coverage, reduce communication costs, and allocate more resources to sensing.
	 Against this background, the main contributions of this paper are summarized as follows:

	 1) We investigate a DP RIS-aided ISAC system that includes a DP ISAC BS, a DP RIS, two radar targets, and communication users with a pair of orthogonally polarized antennas. 
	 This configuration offers enhanced capabilities in jointly optimizing communication and sensing performance compared to conventional SP systems. However, the main challenge in our DP RIS-aided ISAC system is the design of the DP beamforming matrix and the DP RIS elements which are coupled and faced with non-convex constraints.
	 
	 2) To address the coupled optimization variables, we propose a two-stage approach. Initially, an alternating optimization (AO) strategy is applied to decompose the complex problem into two tractable subproblems.
	 For the DP beamforming matrix optimization, we develop a penalty-driven algorithm to handle non-convex signa-to-noise-ratio (SNR) and power constraints. By effectively addressing the non-convex constraints using the penalty-driven algorithm, the optimal solutions to the penalized problem can be obtained using Lagrange duality.
	 Next, the DP RIS phase-shift matrix is optimized using the majorization-minimization (MM) technique, which can obtain the closed-form expression in each iteration.

	 3) Simulation results show that the proposed DP RIS-aided ISAC system  substantially outperforms traditional SP systems in terms of sum rate. Besides, the system performance improves with higher XPD values of DP channels until reaching a saturation point. A notable trade-off can be observed between sensing quality and communication performance, where enhanced sensing requirements result in reduced communication rates. We show the beampattern under different polarization waves and the total beampattern, which validates the system's capability to effectively maintain sensing and communication functionalities.

	The remainder of this paper is organized as follows. Section II introduces the proposed DP system model and formulates the optimization problem for the DP RIS-aided ISAC system. Section III simplifies the proposed problem and optimizes the DP beamforming matrix and the DP RIS reflection coefficients. Section IV shows the results of our simulations. The paper concludes in Section V, where we present our findings.
	
	\textbf{Notation:} 
	Vectors are denoted by bold lowercase letters  and matrices by bold uppercase letters.
	The operations of transpose, conjugate, and Hermitian are  $(\cdot)^{\rm{T}}$, $(\cdot)^{\rm{*}}$, and $(\cdot)^{\rm{H}}$, respectively.
	The space of complex matrices of size $M \times N$ is denoted by $\mathbb{C}^{M \times N}$. The real part of a scalar $x$ is expressed as $\operatorname{Re}(x)$, and the trace of a matrix $\mathbf{A}$ is denoted by $\text{tr}(\mathbf{A})$. 
	The vector $\mathbf{1}_N$ represents an $N\times1$ all-ones vector, and $\mathbf{I}_N$ denotes an $N\times N$ identity matrix.
	The expectation of a random variable is represented by $\mathbb{E}\left[ \cdot \right]$. 
	The $l_2$-norm and Frobenius norm are denoted by $\Vert\cdot\Vert_{2}$ and $\Vert\cdot\Vert_F$, respectively, while $|\cdot|$ represents the absolute value of a scalar.
	A circularly symmetric complex Gaussian random variable with zero mean and variance $\sigma^2$ is denoted by $\mathcal{CN}(0, \sigma^2)$. 
	For a vector $\mathbf{x}$, $\text{diag}(\mathbf{x})$ represents a diagonal matrix with the elements of $\mathbf{x}$ on its main diagonal, while for a matrix $\mathbf{X}$, $\text{vecd}(\mathbf{X})$ denotes a vector containing the diagonal elements of $\mathbf{X}$. 
	The Hadamard (element-wise) product  and the Kronecker product are denoted by \(\odot\) and \(\otimes\), respectively.

	\section{SYSTEM MODEL AND PROBLEM FORMULATION}
			\begin{figure}[bhtp]
			\centering\vspace*{-0.50\baselineskip}
			\includegraphics[width=3.7in]{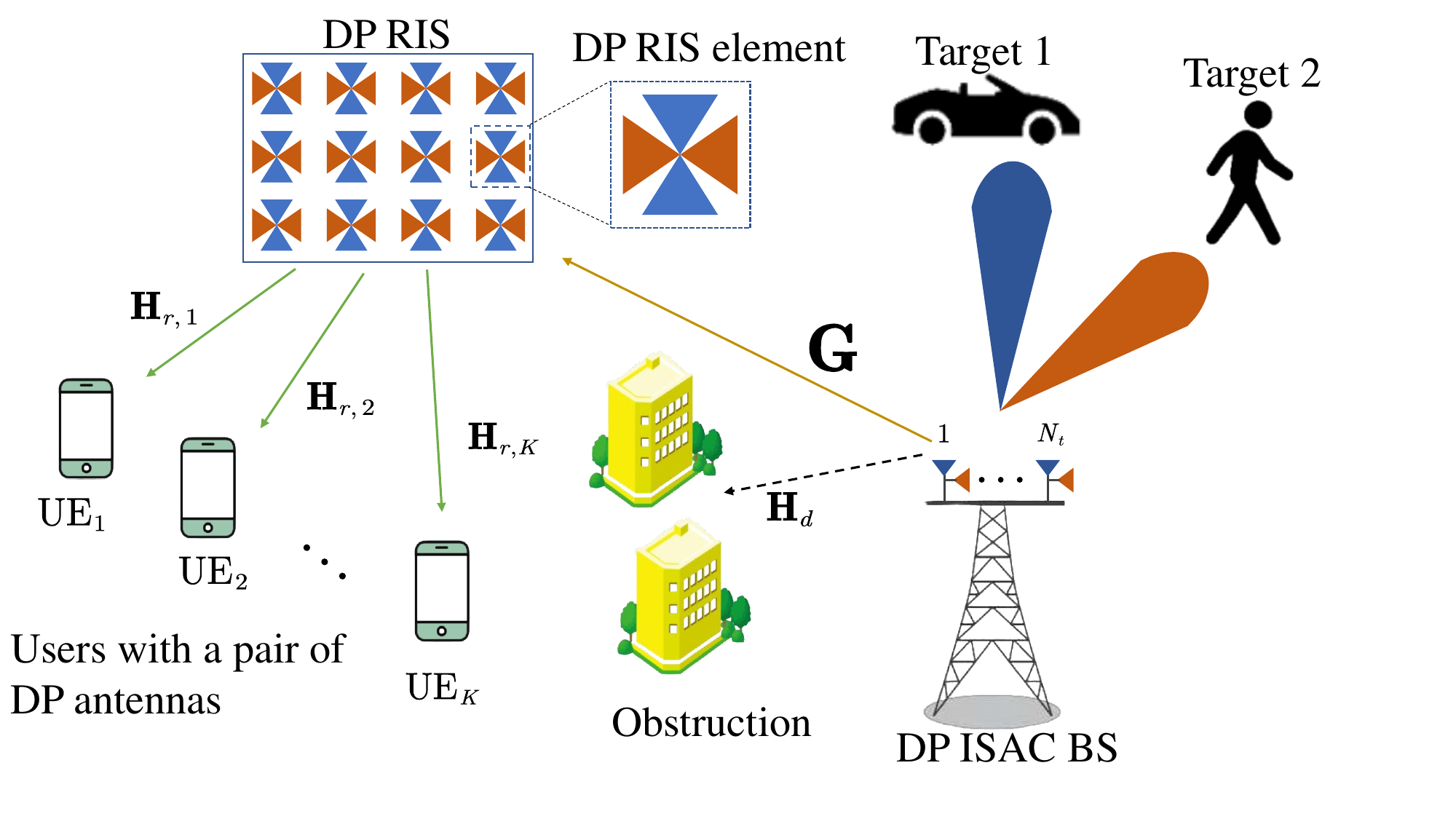}
			\caption{A DP RIS-aided ISAC system.}
			\label{isac_model}
		\end{figure}
	As shown in Fig.~\ref{isac_model}, we consider a DP RIS-aided ISAC system over a DP channel.
	The DFRC BS is equipped with an \(N_t\)-element transmit array and an  $N_r$-element receive array, where each element consists of a pair of DP antennas. 
	Exploiting the potential of the DP architecture, the BS transmits horizontally polarized signals and vertically polarized signals to detect two targets in different orientations, while maintaining communication links with \(K\) user equipments (UEs) through the DP channel	\footnote{The analysis can be extended to other orthogonal  polarization setups, including slanted linear polarizations at $\pm45^\circ$ or circular polarizations rotating in opposite directions.  The conclusions drawn are universally applicable, provided that the spatial correlation is consistent across both polarization types.}.
	A DP RIS with $L$ DP elements arranged in a ULA is mounted on a building facade between BS and users.

	\subsection{DP transmission signal}
In a DP ISAC system, the DP signal transmitted from the DP BS is given by \cite{ozdogan2022massive}
\begin{equation}\label{signalmodel}
	\mathbf{x} = \sum_{k=1}^{K} \mathbf{F}_k \mathbf{s}_k,
\end{equation} 
where \(\mathbf{s}_k =[s_{kv}\ s_{kh}]^{\mathrm{T}} \in \mathbb{C}^{2 \times 1}\) represents the DP transmission symbols intended for the \(K\) users which satisfies $\mathbb{E}\{\mathbf{s}_k\mathbf{s}_k^{\mathrm{H}}=\mathbf{I}_2\}$ and $\mathbb{E}[\mathbf{s}_k\mathbf{s}_j^{\mathrm{H}}] = \mathbf{0}$, for $k \neq j$.
Here, we use the labels `v' and `h' to represent vertical polarization and horizontal polarization for ease, respectively. $\mathbf{F}_k=[\mathbf{f}_{kv}\ \mathbf{f}_{kh}]\in\mathbb{C}^{2N_t\times2}$ is the downlink precoding matrix for the DP transmit array.
 Furthermore, considering the hardware limitations and system requirements, the total transmit power needs to satisfy
\begin{equation}
	\mathbb{E}\left[\|\mathbf{x}\|^2\right] = \sum_{k=1}^K \text{tr}\left(\mathbf{F}_k\mathbf{F}_k^{\mathrm{H}}\right) \leq P_0,
\end{equation}
where $P_0$ represents the maximum permissible transmit power of the system.

\subsection{DP channel model}
As illustrated in Fig.~\ref{isac_model}, the downlink transmission in our DP ISAC system comprises two primary paths: a direct link between the BS and the user, and a reflected link assisted by the DP RIS. The total fast fading DP channel from the DP BS to UE $k$ can be expressed as
\begin{equation}\label{dpchannel}
	\mathbf{H}_k = \mathbf{H}_{d,k}+\mathbf{H}_{r,k} \mathbf{\Phi} \mathbf{G}  ,
\end{equation}
where $\mathbf{H}_{d,k} \in \mathbb{C}^{2\times2N_t}$ represents the fast fading channel between the DP BS and user $k$, $\mathbf{H}_{r,k} \in \mathbb{C}^{2\times2L}$ denotes the fast fading channel between the DP RIS and user $k$, $\mathbf{G} \in \mathbb{C}^{2L \times 2N_t}$ characterizes the fast fading channel from the DP BS to the DP RIS, $\mathbf{\Phi} \in \mathbb{C}^{2L \times 2L}$ represents phase-shift matrix of the DP RIS.

In the following, we introduce the polarization-specific channel representations of the fast fading channel between the BS and users as an example. Specifically, for UE $k$ with a pair of DP antennas, the channel $\mathbf{H}_{d,k}$ between the BS and UE $k$ can be partitioned as
\begin{equation}
	\mathbf{H}_{d,k} = \begin{bmatrix}
		\mathbf{h}_{d,k,vv} & \mathbf{h}_{d,k,vh} \\
		\mathbf{h}_{d,k,hv} & \mathbf{h}_{d,k,hh}\
	\end{bmatrix},
\end{equation}
where $\mathbf{h}_{d,k,pq} \in \mathbb{C}^{1\times N_t}$ represents the channel component from BS antennas with polarization $q$ to UE $k$'s antenna with polarization $p$, where $\{p,q\}\in\{v,h\}$ represent the respective antenna polarization. 
Due to the presence of multiple obstacles between the BS and the UE in practical scenarios, there is only non-line-of-sight (NLoS) link. 
 Therefore, each element  of $\mathbf{h}_{d,k,pq}$, denoted by $\mathbf{h}_{d,k,pq}(n)$, $n=1,\cdots,N_t$, follows a zero-mean independent and identically distributed (i.i.d.) complex Gaussian distribution.

A key characteristic of DP systems is polarization conversion, which introduces imperfections in signal transmission. To quantify the system's ability to maintain polarization purity, the XPD metric of the DP channel is defined as
\begin{align}\label{XPD}
	\mathrm{XPD} = \frac{\mathbb{E}\left\{ \lVert \mathbf{h}_{d,k,vv} \rVert^2 \right\}}{\mathbb{E}\left\{ \lVert \mathbf{h}_{d,k,hv} \rVert^2 \right\}} = \frac{\mathbb{E}\left\{ \lVert \mathbf{h}_{d,k,hh} \rVert^2 \right\}}{\mathbb{E}\left\{ \lVert \mathbf{h}_{d,k,hv} \rVert^2 \right\}},
\end{align}
where the cross-polarized components $\mathbf{h}_{d,k,hv}(n)$ and $\mathbf{h}_{d,k,vh}(n)$ exhibit a variance of $\beta_1^{\text{NLoS}}$ ($0 < \beta_1^{\text{NLoS}} \leq 1$), representing the degree of polarization coupling between orthogonal states. Following established channel modeling principles\cite{coldrey2008modeling}, the co-polarized components $\mathbf{h}_{d,k,vv}(n)$ and $\mathbf{h}_{d,k,hh}(n)$ are characterized by a variance of $1-\beta_1^{\text{NLoS}}$ to ensure power conservation in the channel. Under assumptions, the following statistical properties are established
\begin{equation}
	\begin{aligned}
		\mathbb{E}\{|\mathbf{h}_{d,k,vv}(n)|^2\} = \mathbb{E}\{|\mathbf{h}_{d,k,hh}(n)|^2\} &= 1-\beta_1^{\text{NLoS}}, \\
		\mathbb{E}\{|\mathbf{h}_{d,k,vh}(n)|^2\} = \mathbb{E}\{|\mathbf{h}_{d,k,hv}(n)|^2\} &= \beta_1^{\text{NLoS}},
	\end{aligned}
\end{equation}
so XPD equals $(1-\beta_1^{\text{NLoS}})/\beta_1^{\text{NLoS}}$ according to \eqref{XPD}. Furthermore, considering the statistical independence of polarization components measured in\cite{Asplund2007}, we have
\begin{equation}
	\begin{aligned}
		\mathbb{E}\{\mathbf{h}_{d,k,vv}(n)\mathbf{h}_{d,k,hv}^*(n)\} = \mathbb{E}\{\mathbf{h}_{d,k,vh}(n)\mathbf{h}_{d,k,hh}^*(n)\} &= 0, \\
		\mathbb{E}\{\mathbf{h}_{d,k,vv}(n)\mathbf{h}_{d,k,hh}^*(n)\} = \mathbb{E}\{\mathbf{h}_{d,k,vh}(n)\mathbf{h}_{d,k,hv}^*(n)\} &= 0,
	\end{aligned}
\end{equation}
which means that the XPC is approximately equal to zero.

The DP RIS is carefully positioned to establish deterministic line-of-sight (LoS) channels with both the BS and the user.
The BS-to-RIS channel $\mathbf{G}$ experiences Rician fading. The LoS and NLoS components of $\mathbf{G}$ exhibit XPD ratios of $(1-\beta_2^{\text{LoS}})/\beta_2^{\text{LoS}}$ and $(1-\beta_2^{\text{NLoS}})/\beta_2^{\text{NLoS}}$, respectively. Channel matrix $\mathbf{G}$ can be decomposed into polarization-specific submatrices as
\begin{equation}
	\mathbf{G} = \begin{bmatrix}
		\mathbf{G}_{vv} & \mathbf{G}_{vh} \\
		\mathbf{G}_{hv} & \mathbf{G}_{hh}\
	\end{bmatrix},
\end{equation}
where $\mathbf{G}_{pq} \in \mathbb{C}^{L\times N_t}$ is the sub-channel component from $q$-polarization antennas of DP BS to $p$-polarization elements of DP RIS. These sub-channels are characterized by
\begin{equation}\label{Gchannel}
	\begin{aligned}
		\mathbf{G}_{vv} &= \omega_{\text{L}}\sqrt{1-\beta_2^{\text{LoS}}}\hat{\mathbf{G}}^{\text{LoS}}+\omega_{\text{N}}\sqrt{1-\beta_2^{\text{NLoS}}}\hat{\mathbf{G}}^{\text{NLoS}}, \\
		\mathbf{G}_{vh} &= \omega_{\text{L}}\sqrt{\beta_2^{\text{LoS}}}\hat{\mathbf{G}}^{\text{LoS}}+\omega_{\text{N}}\sqrt{\beta_2^{\text{NLoS}}}\hat{\mathbf{G}}^{\text{NLoS}},  \\
		\mathbf{G}_{hv} &= \omega_{\text{L}}\sqrt{\beta_2^{\text{LoS}}}\hat{\mathbf{G}}^{\text{LoS}}+\omega_{\text{N}}\sqrt{\beta_2^{\text{NLoS}}}\hat{\mathbf{G}}^{\text{NLoS}},  \\
		\mathbf{G}_{hh} &= \omega_{\text{L}}\sqrt{1-\beta_2^{\text{LoS}}}\hat{\mathbf{G}}^{\text{LoS}}+\omega_{\text{N}}\sqrt{1-\beta_2^{\text{NLoS}}}\hat{\mathbf{G}}^{\text{NLoS}},
	\end{aligned}
\end{equation}
where $\omega_{\text{L}}=\sqrt{\frac{\omega}{\omega+1}}$, $\omega_{\text{N}}=\sqrt{\frac{1}{\omega+1}}$, and $\omega$ is the Rician factor. Here, $\hat{\mathbf{G}}^{\text{LoS}} \in \mathbb{C}^{L \times N_t}$ and $\hat{\mathbf{G}}^{\text{NLoS}} \in \mathbb{C}^{L \times N_t}$ denote the shared channel components for LoS and NLoS paths.

	Similarly, for the channel $\mathbf{H}_{r,k}$ from the DP RIS to UE $k$,
	we partition $\mathbf{H}_{r,k}$ as
	\begin{equation}
		\mathbf{H}_{r,k} = \begin{bmatrix}
			\mathbf{h}_{r,k,vv} & \mathbf{h}_{r,k,vh} \\
			\mathbf{h}_{r,k,hv} & \mathbf{h}_{r,k,hh}\
		\end{bmatrix},
	\end{equation}
	where $\mathbf{h}_{r,k,pq} \in \mathbb{C}^{1\times L}$ represents the channel component from DP RIS elements with polarization $q$ to UE $k$'s antenna with polarization $p$.
	These respective sub-channels are
	\begin{equation}\label{hchannel}
		\begin{aligned}
			\mathbf{h}_{r,k,vv} &= \omega_{\text{L}}\sqrt{1-\beta_3^{\text{LoS}}}\hat{\mathbf{h}}_{r,k}^{\text{LoS}}+\omega_{\text{N}}\sqrt{1-\beta_3^{\text{NLoS}}}\hat{\mathbf{h}}_{r,k}^{\text{NLoS}}, \\
			\mathbf{h}_{r,k,vh} &= \omega_{\text{L}}\sqrt{\beta_3^{\text{LoS}}}\hat{\mathbf{h}}_{r,k}^{\text{LoS}}+\omega_{\text{N}}\sqrt{\beta_3^{\text{NLoS}}}\hat{\mathbf{h}}_{r,k}^{\text{NLoS}},  \\
			\mathbf{h}_{r,k,hv} &= \omega_{\text{L}}\sqrt{\beta_3^{\text{LoS}}}\hat{\mathbf{h}}_{r,k}^{\text{LoS}}+\omega_{\text{N}}\sqrt{\beta_3^{\text{NLoS}}}\hat{\mathbf{h}}_{r,k}^{\text{NLoS}},  \\
			\mathbf{h}_{r,k,hh} &= \omega_{\text{L}}\sqrt{1-\beta_3^{\text{LoS}}}\hat{\mathbf{h}}_{r,k}^{\text{LoS}}+\omega_{\text{N}}\sqrt{1-\beta_3^{\text{NLoS}}}\hat{\mathbf{h}}_{r,k}^{\text{NLoS}},
		\end{aligned}
	\end{equation}
	where $\hat{\mathbf{h}}_{r,k}^{\text{LoS}}$, $\hat{\mathbf{h}}_{r,k}^{\text{NLoS}}$ represent the shared fading components for LoS and NLoS links, and $\beta_3^{\text{LoS}}$, $\beta_3^{\text{NLoS}} \in [0,1]$ denote the corresponding XPD parameters, similar to those in the BS-RIS channel.

		\subsection{DP RIS}
		
		A DP RIS consists of \(L\) reflecting elements capable of performing polarization beam splitting, independent control of impinging polarizations, and polarization conversion. The phase-shift matrix of the DP RIS is defined as
		\begin{align}
			\mathbf{\Phi} = \begin{bmatrix}
				\mathbf{\Phi}_{vv} & \mathbf{\Phi}_{vh} \\
				\mathbf{\Phi}_{hv} & \mathbf{\Phi}_{hh}
			\end{bmatrix} \in \mathbb{C}^{2L \times 2L},
		\end{align}
		where \(\mathbf{\Phi}_{pq} = \mathrm{diag}\left\{ \phi_{1}^{pq}, \ldots, \phi_{L}^{pq} \right\}\). The superscripts \(\{pq\}\) indicate the transition from \(p\)-polarization to \(q\)-polarization through the dual-polarized RIS. The phase shift \(\phi_{n}^{pq} = a_n^{pq} \exp(j\theta_n^{pq})\), where \(a_n^{pq}\) is the amplitude reflection coefficient and \(\theta_n^{pq} \in [0, 2\pi)\) denotes the phase shift applied to the incident waves in polarization state \(\{pq\}\). 
		
		Notably, previous studies often neglected the $vh$ and $hv$ components in RIS modeling, assuming perfect polarization isolation for analytical tractability. To provide a comprehensive understanding of the DP RIS, we incorporate these cross-polarization components in our model considering that the polarization coupling
		effects are non-negligible in practical systems.

		\section{PROBLEM FORMULATION}
		\subsection{Communication performance metric}
		 Assuming perfect channel state information (CSI) is available at the BS, the received signal at UE $k$, denoted by $\mathbf{y}_k \in \mathbb{C}^{2\times1}$, can be expressed as
		\begin{equation}\label{receivesignal}
			\mathbf{y}_k =  \mathbf{H}_k \mathbf{x}  + \mathbf{n}_k,
		\end{equation}
		where $\mathbf{n}_k \in \mathbb{C}^{2\times1}$ is the noise vector following complex Gaussian distribution $\mathcal{CN}(\mathbf{0},\sigma_{k}^2\mathbf{I}_{2})$, with $\sigma_k^2$ denoting the noise power spectral density at user $k$.
		Substituting \eqref{signalmodel} into \eqref{receivesignal}, we can rewrite $\mathbf{y}_k$ as
		\begin{equation}
			\mathbf{y}_k = \mathbf{H}_k \sum_{i=1}^K \mathbf{F}_i\mathbf{s}_i + \mathbf{n}_k.
		\end{equation}
		Let $\mathbf{J}_k = \sigma^2\mathbf{I}_{2} + \sum_{i=1,i\neq k}^K \mathbf{H}_k\mathbf{F}_i\mathbf{F}_i^{\mathrm{H}}\mathbf{H}_k^{\mathrm{H}}$ denote the interference-plus-noise covariance matrix at user $k$. The achievable sum rate (in nat/s/Hz) of the considered system can be expressed as
		\begin{equation}\label{sumrate}
			R = \sum_{k=1}^K \log_2|\mathbf{I}_{2} + \mathbf{H}_k\mathbf{F}_k\mathbf{F}_k^{\mathrm{H}}\mathbf{H}_k^{\mathrm{H}}\mathbf{J}_k^{-1}|.
		\end{equation}
		
		\subsection{Sensing performance metric}
	For sensing, we propose a sensing scheme that orthogonally polarized waves simultaneously illuminate targets in different directions to exploit the potential of DP array. Compared to SP arrays, this approach enables concurrent multi-target detection while maintaining the same sensing performance. Furthermore, it can obtain polarimetric scattering characteristics of targets, facilitating enhanced target recognition and classification capabilities.
		 
	 For the sensing functionality, the received echo signal at the radar receiver for target 1 and target 2 in respective directions can be modeled as
		 \begin{equation}
		 	\begin{aligned}
		 		&\mathbf{y}_{1} =\mathbf{A}_1\mathbf{x}_1 + \mathbf{z}_{1},\\
		 		&\mathbf{y}_{2} =\mathbf{A}_2\mathbf{x}_2 + \mathbf{z}_{2},\
		 	\end{aligned}
		 \end{equation}
		 where $\mathbf{x}_1=\mathbf{E}_1\mathbf{x}$ and $\mathbf{x}_2=\mathbf{E}_2\mathbf{x}$ are the vertical and horizontal part of transmit signal $\mathbf{x}$, where 
		 \begin{align}\label{EvEh}
		 	\mathbf{E}_1=[\mathbf{I}_{N_t}\ \mathbf{0}_{N_t}],\ \mathbf{E}_2=[\mathbf{0}_{N_t}\ \mathbf{1}_{N_t}].
		 \end{align}
		 $\mathbf{z}_{1},\mathbf{z}_{2}$ represents the AWGN satisfying $\mathcal{CN}(0,\sigma_{\text{r}}^2\mathbf{I}_{N_{\text{r}}})$ at the radar receiver with the variance $\sigma_{\text{r}}^2$. The matrix $\mathbf{A}_1,\mathbf{A}_2 \in \mathbb{C}^{N_r\times N_t}$ denotes the target response matrix for target 1 and target 2. The point target response matrix can be defined as
		 \begin{equation}
		 	\begin{aligned}
		 		\mathbf{A}_1 = \mathbf{a}_{N_r}(\theta_1)\mathbf{a}_{N_t}^\mathrm{H}(\theta_1),\\
		 	\mathbf{A}_2 = \mathbf{a}_{N_r}(\theta_2)\mathbf{a}_{N_t}^\mathrm{H}(\theta_2),
		 	\end{aligned}
		 \end{equation}
		 where $\theta_1$, $\theta_2$ represent the direction of target 1 and target 2 with respect to (w.r.t.) the BS. The vector $\mathbf{a}_{N}(\theta)$ is the steering vector associated with the BS, defined by
		 \begin{equation}\label{steervector}
		 	\mathbf{a}_N(\theta) = \left[1,\cdots,e^{-j2\pi d(N-1)\sin\theta/\lambda}\right]^\mathrm{T},
		 \end{equation}
		 where $d$ and $\lambda$ denote the element spacing and the signal wavelength, and $N$ is the array size.
		 
		 Take target 1 as an example, the output received by the radar after beamforming can be expressed as
		 \begin{equation}
		 	r_1 = \mathbf{w}_{1}^\mathrm{H}\mathbf{y}_{1} = \eta_1\mathbf{w}_{1}^\mathrm{H}\mathbf{A}_1\mathbf{x}_1 + \mathbf{w}_{1}^\mathrm{H}\mathbf{z}_{1},
		 \end{equation}
		 where $\eta_1$ denotes the channel gain for target 1 and
		  $\mathbf{w}_{1} \in \mathbb{C}^{{N_r}\times 1}$ represents the receive beamforming vector for target 1. To maximize the radar SNR, the minimum variance distortionless response (MVDR) algorithm is utilized, which gives the optimal $\mathbf{w}_{1}^*$ as
		 \begin{equation}
		 	\mathbf{w}_{1}^* = \arg\max_{\mathbf{w}_{1}} \frac{|\mathbf{w}_{1}^\mathrm{H}\mathbf{V}_1\mathbf{x}_1|^2}{\mathbf{w}_{1}^\mathrm{H}\mathbf{w}_{1}} = \beta_1\mathbf{V}_1\mathbf{x}_1,
		 \end{equation}
		 where $\beta_1$ represents a constant and $\mathbf{V}_1 \triangleq \eta_1\mathbf{A}_1$. Thus, the corresponding radar SNR can be computed as
		 \begin{equation}
		 	\begin{aligned}
		 		\gamma_1 &= \mathbb{E}\left[\frac{|\mathbf{w}_{1}^\mathrm{H}\mathbf{V}_1\mathbf{x}_1|^2}{\sigma_{\text{1}}^2\mathbf{w}_{1}^\mathrm{H}\mathbf{w}_{1}}\right]\\ 
		 		&\stackrel{(a)}{=} \mathbb{E}[\mathbf{x}_1^\mathrm{H}\mathbf{V}_1^\mathrm{H}\mathbf{V}_1\mathbf{x}]/\sigma_{\text{r}}^2\\
		 		&\stackrel{(b)}{=} \sum_{k=1}^K \text{tr}((\mathbf{V}_1\mathbf{E}_1)^\mathrm{H}(\mathbf{V}_1\mathbf{E}_1)\mathbf{F}_k\mathbf{F}_k^\mathrm{H})/\sigma_{\text{r}}^2,
		 	\end{aligned}
		 \end{equation}
		 where (a) corresponds to the determination of the optimal receive beamforming vector $\mathbf{w}_{1}^*$, and (b) results from the mathematical expectation $\mathbb{E}[\mathbf{x}_1\mathbf{x}_1^\mathrm{H}] = \sum_{k=1}^K \mathbf{E}_1\mathbf{F}_k\mathbf{F}_k^\mathrm{H}\mathbf{E}_1^{\mathrm{H}}$.
		 Similarly, for target 2, the radar SNR is
		 \begin{align}
		 	\gamma_2= \sum_{k=1}^K \text{tr}((\mathbf{V}_2\mathbf{E}_2)^\mathrm{H}(\mathbf{V}_2\mathbf{E}_2)\mathbf{F}_k\mathbf{F}_k^\mathrm{H})/\sigma_{\text{r}}^2,
		 \end{align}
		 where $\mathbf{V}_2 \triangleq \eta_2\mathbf{A}_2$, and $\eta_2$ is the channel gain for target 2.

		\subsection{Optimization Model}
		In this paper, we aim to jointly optimize the DP transmit beamforming matrix $\mathbf{F}$ and the DP phase-shift matrix $\mathbf{\Phi}$ to maximize the system sum rate while ensuring effective dual-target sensing performance. The optimization problem can be formulated as
		\begin{subequations}\label{optimization_problem}
			\begin{align}
				\max_{\mathbf{F}, \mathbf{\Phi}} \quad & R \label{P1_a} \\
				\text{s.t.} \quad & \gamma_1 \geq \gamma_{\text{1,th}}, \label{P1_radar1} \\
				& \gamma_2 \geq \gamma_{\text{2,th}}, \label{P1_radar2} \\
				& \sum_{k=1}^K\|\mathbf{F}_k\|_{F}^2 \leq P_0, \label{P1_c} \\ 
				& |\phi_{n,pq}| = 1, \quad \forall n, \{p,q\} \in \{v,h\}, \label{P1_d}
			\end{align}
		\end{subequations}
		where $\gamma_1$ and $\gamma_2$ represent the sensing SNR for targets 1 and 2, respectively, $\gamma_{\text{1,th}}$ and $\gamma_{\text{2,th}}$ denote the required SNR threshold for reliable target detection, and $P_0$ is the maximum transmit power.
		
		The main challenge in solving this optimization problem stems from the complex coupling between the transmit beamforming matrix $\mathbf{F}$ and the phase-shift matrix $\mathbf{\Phi}$. These variables are coupled in both the objective function and the dual-target sensing constraints (\ref{P1_radar1})-(\ref{P1_radar2}). Moreover, the constant modulus constraint (\ref{P1_d}) for RIS phase shifts introduces additional non-convexity, making the problem particularly challenging.

		\section{PROPOSED ALGORITHM}
		In this section, we first reformulate Problem \eqref{optimization_problem} into an equivalent form by exploiting the relationship between the sum rate and the weighted minimum mean-square error (WMMSE) principle. We then employ the alternating optimization (AO) method to tackle the reformulated problem, considering the inherent coupling among optimization variables. This method enables us to decompose the complex problem into two tractable subproblems, each focusing on specific variable optimization, thereby enhancing the overall solution efficiency. For the DP transmit beamforming matrix optimization subproblem, we develop a penalty-based algorithm. Meanwhile, we adopt the MM algorithm to handle the constant module constraint on the DP phase-shift matrix. This comprehensive approach effectively addresses the coupling between the DP beamforming matrix $\mathbf{F}$ and the phase-shift matrix $\mathbf{\Phi}$ while managing the non-convex constraints, ultimately improving the system performance.

	\subsection{Reformulation of the Original Problem}
	
Without loss of generality, we consider UE $k$ with a pair of orthogonally polarized antennas and derive the problem reformulation.
A linear receive filter matrix $\mathbf{U}_k \in \mathbb{C}^{2 \times 1}$ is introduced to estimate the signal vector $\hat{\mathbf{d}}_k$ for UE $k$ 
	by
	\begin{equation}
		\hat{\mathbf{d}}_k = \mathbf{U}_k^{\mathrm{H}}\mathbf{y}_k, \forall k.
		\label{eq:13}
	\end{equation}
	The corresponding mean-square error (MSE) matrix for user $k$ is derived as
	\begin{equation}
		\begin{aligned}
			\mathbf{E}_k &= \mathbb{E}\left[\left(\hat{\mathbf{d}}_k - \mathbf{d}_k\right)\left(\hat{\mathbf{d}}_k - \mathbf{d}_k\right)^{\mathrm{H}}\right] \\
			&= \mathbf{U}_k^{\mathrm{H}}\mathbf{H}_k\left(\sum_{i=1}^K \mathbf{F}_i\mathbf{F}_i^{\mathrm{H}}\right)\mathbf{H}_k^{\mathrm{H}}\mathbf{U}_k - \mathbf{U}_k^{\mathrm{H}}\mathbf{H}_k\mathbf{F}_k \\
			&\quad -\mathbf{F}_k^{\mathrm{H}}\mathbf{H}_k^{\mathrm{H}}\mathbf{U}_k + \sigma_k^2\mathbf{U}_k^{\mathrm{H}}\mathbf{U}_k + \mathbf{I}_{2}.
			\label{eq:14}
		\end{aligned}
	\end{equation}
	Then, the expression in \eqref{sumrate} can be rewritten as
	\begin{equation}
		E = \sum_{k=1}^K \left(\log|\mathbf{W}_k| - \text{tr}(\mathbf{W}_k\mathbf{E}_k) + 2\right),
		\label{sum_MSE}
	\end{equation}	
where $\mathbf{W}_k \in \mathbb{C}^{2 \times 2}$ serves as the auxiliary weight matrix for the $k$-th user, and the constant term `2' corresponds to the dimension of the data stream due to DP transmission.
	
	After removing the constant terms from \eqref{sum_MSE}, Problem \eqref{optimization_problem} is converted to
	\begin{subequations}
		\begin{align}
			\min_{\{\mathbf{F}_k,\mathbf{W}_k,\mathbf{U}_k\},\boldsymbol{\Phi}} &\sum_{k=1}^K \left(\text{tr}(\mathbf{W}_k\mathbf{E}_k) - \log|\mathbf{W}_k|\right) \label{eq:16a} \\
			\text{s.t.} \quad & \eqref{P1_radar1},\eqref{P1_radar2},\eqref{P1_c},\eqref{P1_d}. \label{eq:16b}
		\end{align}\label{eq:16}
	\end{subequations}
	The joint optimization of variables in Problem \eqref{eq:16} presents  computational challenges. To develop an efficient solution approach, we implement an AO framework that sequentially updates different variable groups. Specifically for Problem \eqref{eq:16}, we observe that variables $\mathbf{U}_k$ and $\mathbf{W}_k$ only appear in the objective function. By taking the derivative of $E$ concerning $\mathbf{U}_k$ and setting it to zero while keeping other variables fixed, we obtain

	\begin{equation}
		\mathbf{U}_k^{\text{opt}} = \left(\mathbf{H}_k\sum_{i=1}^K \mathbf{F}_i\mathbf{F}_i^{\mathrm{H}}\mathbf{H}_k^H + \sigma_k^2\mathbf{I}_{2}\right)^{-1}\mathbf{H}_k\mathbf{F}_k.
		\label{eq:17}
	\end{equation}
	
	In the same way, with other variables remaining constant, the optimal $\mathbf{W}_k$ is determined as
	\begin{equation}
		\mathbf{W}_k^{\text{opt}} = \mathbf{E}_k^{-1}.
		\label{eq:18}
	\end{equation}

		\subsection{Optimization of  the DP Phase-Shift Matrix $\mathbf{\Phi}$}
		In the following, our primary focus lies in optimizing matrix $\boldsymbol{\Phi}$ while keeping $\{\mathbf{F}_k\}$, $\{\mathbf{W}_k\}$, and $\{\mathbf{U}_k\}$ fixed. 
		For notational convenience, we define the vectorized forms of $\mathbf{\Phi}$ as follows: $\boldsymbol{\phi}_{vv} = \text{vecd}(\mathbf{\Phi}_{vv})$, $\boldsymbol{\phi}_{vh} = \text{vecd}(\mathbf{\Phi}_{vh})$, $\boldsymbol{\phi}_{hv} = \text{vecd}(\mathbf{\Phi}_{hv})$, $\boldsymbol{\phi}_{hh} = \text{vecd}(\mathbf{\Phi}_{hh})$, which are aggregated into a single vector: ${\bm{\phi}}=[\bm{\phi}_{vv},\bm{\phi}_{vh},\bm{\phi}_{hv},\bm{\phi}_{hh}]^{\mathrm{T}}$. 
		By substituting the expression for $\mathbf{E}_k$ into \eqref{sum_MSE} and eliminating constant terms, we can formulate the phase-shift optimization problem as follows
		\begin{subequations}
			\begin{align}
				\min_{\boldsymbol{\Phi}} \quad & -\sum_{k=1}^K \text{tr}\left(\mathbf{W}_k\mathbf{U}_k^{\mathrm{H}}\mathbf{H}_k\mathbf{F}_k\right) \nonumber \\
				& +\sum_{k=1}^K \text{tr}\left(\mathbf{W}_k\mathbf{U}_k^{\mathrm{H}}\mathbf{H}_k\sum_{i=1}^K \mathbf{F}_i\mathbf{F}_i^{\mathrm{H}}\mathbf{H}_k^{\mathrm{H}}\mathbf{U}_k\right) \nonumber \\
				& -\sum_{k=1}^K \text{tr}\left(\mathbf{W}_k\mathbf{F}_k^{\mathrm{H}}\mathbf{H}_k^{\mathrm{H}}\mathbf{U}_k\right) \label{eq:40a} \\
				\text{s.t.} \quad &  \eqref{P1_d}. \label{eq:40b}
			\end{align}\label{eq:40}
		\end{subequations}
		Let  $\mathbf{F} = \sum_{k=1}^K \mathbf{F}_k\mathbf{F}_k^{\mathrm{H}}$, by substituting $\mathbf{H}_k = \mathbf{H}_{\text{d},k} + \mathbf{H}_{\text{r},k}\boldsymbol{\Phi}\mathbf{G}$ into the expression $\mathbf{W}_k\mathbf{U}_k^{\mathrm{H}}\mathbf{H}_k\sum_{i=1}^K \mathbf{F}_i\mathbf{F}_i^{\mathrm{H}}\mathbf{H}_k^{\mathrm{H}}\mathbf{U}_k$, we obtain
		\begin{equation}
			\begin{aligned}
				\mathbf{W}_k\mathbf{U}_k^{\mathrm{H}}\mathbf{H}_k\sum_{i=1}^K& \mathbf{F}_i\mathbf{F}_i^{\mathrm{H}}\mathbf{H}_k^{\mathrm{H}}\mathbf{U}_k =  \mathbf{W}_k\mathbf{U}_k^{\mathrm{H}}\mathbf{H}_{\text{d},k}\mathbf{F}\mathbf{H}_{\text{d},k}^{\mathrm{H}}\mathbf{U}_k \\
				& +\mathbf{W}_k\mathbf{U}_k^{\mathrm{H}}\mathbf{H}_{r,k}\boldsymbol{\Phi}\mathbf{G}\mathbf{F}\mathbf{G}^{\mathrm{H}}\boldsymbol{\Phi}\mathbf{H}_{r,k}^{\mathrm{H}}\mathbf{U}_k \\
				& +\mathbf{W}_k\mathbf{U}_k^{\mathrm{H}}\mathbf{H}_{r,k}\boldsymbol{\Phi}\mathbf{G}\mathbf{F}\mathbf{H}_{\text{d},k}^{\mathrm{H}}\mathbf{U}_k \\
				& +\mathbf{W}_k\mathbf{U}_k^{\mathrm{H}}\mathbf{H}_{\text{d},k}\mathbf{F}\mathbf{G}^{\mathrm{H}}\boldsymbol{\Phi}^{\mathrm{H}}\mathbf{H}_{r,k}^{\mathrm{H}}\mathbf{U}_k
			\end{aligned}
			\label{eq:41}
		\end{equation}
		
		and
		\begin{equation}
			\mathbf{W}_k\mathbf{U}_k^{\mathrm{H}}\mathbf{H}_k\mathbf{F}_k = \mathbf{W}_k\mathbf{U}_k^{\mathrm{H}}\mathbf{H}_{r,k}\boldsymbol{\Phi}\mathbf{G}\mathbf{F}_k + \mathbf{W}_k\mathbf{U}_k^{\mathrm{H}}\mathbf{H}_{\text{d},k}\mathbf{F}_k.
			\label{eq:42}
		\end{equation}
		Then, we define $\mathbf{F}_k = \mathbf{H}_{r,k}^{\mathrm{H}}\mathbf{U}_k\mathbf{W}_k\mathbf{U}_k^{\mathrm{H}}\mathbf{H}_{r,k}$, $\mathbf{C} = \mathbf{G}\mathbf{F}\mathbf{G}^{\mathrm{H}}$ and $\mathbf{M}_k^{\mathrm{H}} = \mathbf{G}\mathbf{F}\mathbf{H}_{\text{d},k}^{\mathrm{H}}\mathbf{U}_k\mathbf{W}_k\mathbf{U}_k^{\mathrm{H}}\mathbf{H}_{r,k}$, so
		\begin{equation}
			\begin{aligned}
				\text{tr}\left(\mathbf{W}_k\mathbf{U}_k^{\mathrm{H}}\mathbf{H}_k\sum_{i=1}^K \mathbf{F}_i\mathbf{F}_i^{\mathrm{H}}\mathbf{H}_k^{\mathrm{H}}\mathbf{U}_k\right) = \text{tr}\left(\mathbf{F}_k\boldsymbol{\Phi}\mathbf{C}\boldsymbol{\Phi}^{\mathrm{H}}\right) +&\\ \text{tr}\left(\mathbf{M}_k\boldsymbol{\Phi}^{\mathrm{H}}\right) +
				\text{tr}\left(\mathbf{M}_k^{\mathrm{H}}\boldsymbol{\Phi}\right) 
				 + \text{const}_1,&\
			\end{aligned}
			\label{eq:43}
		\end{equation}
		and $\text{const}_1$ 
		\begin{align}
			\text{const}_1 = 	\text{tr}\left(\mathbf{W}_k\mathbf{U}_k^{\mathrm{H}}\mathbf{H}_{d,k}\sum_{k=i}^K \mathbf{F}_i\mathbf{F}_i^{\mathrm{H}}\mathbf{H}_{d,k}^{\mathrm{H}}\mathbf{U}_k\right), 
		\end{align}
		which is not dependent on $\mathbf{\Phi}$.
		
		Similarly, defining $\mathbf{N}_k^{\mathrm{H}} = \mathbf{G}\mathbf{N}_k\mathbf{W}_k\mathbf{U}_k^{\mathrm{H}}\mathbf{H}_{\text{ru},k}$, we have
		\begin{equation}
			\text{tr}\left(\mathbf{W}_k\mathbf{U}_k^{\mathrm{H}}\mathbf{H}_k\mathbf{F}_k\right) = \text{tr}\left(\boldsymbol{\Phi}\mathbf{N}_k^{\mathrm{H}}\right) +
			\text{const}_2,
			\label{eq:44}
		\end{equation}
		where $\text{const}_2$ is a constant term that does not depend on $\boldsymbol{\Phi}$.
		
		From \eqref{eq:43} and \eqref{eq:44}, the objective function \eqref{eq:40a} can be derived as
		\begin{equation}
			\text{tr}\left(\mathbf{F}\boldsymbol{\Phi}\mathbf{C}\boldsymbol{\Phi}^{\mathrm{H}}\right) + \text{tr}\left(\mathbf{\Phi}\mathbf{P}\right) + \text{tr}\left(\mathbf{P}^{\mathrm{H}}\boldsymbol{\Phi}^{\mathrm{H}}\right),
			\label{eq:45}
		\end{equation}
		where $\mathbf{F} = \sum_{k=1}^K \mathbf{F}_k$ and $\mathbf{P} = \sum_{k=1}^K \left(\mathbf{M}_k^{\mathrm{H}} - \mathbf{N}_k^{\mathrm{H}}\right)$. 
		
		For the simplification of $\text{tr}\left(\mathbf{F}\boldsymbol{\Phi}\mathbf{C}\boldsymbol{\Phi}^{\mathrm{H}}\right)$,
		let \(\mathbf{\Xi}\) be a \(4 \times 4\) block matrix consisting of \(N_t \times N_t\) blocks. The matrices $\mathbf{F}$ and $\mathbf{C}$ are structured as $2 \times 2$ block matrices, where each block has dimensions $N_t \times N_t$:
		\begin{align}
			\mathbf{F} = \begin{bmatrix} 
				\mathbf{F}_{00} & \mathbf{F}_{01} \\
				\mathbf{F}_{10} & \mathbf{F}_{11} 
			\end{bmatrix}, \quad
			\mathbf{C} = \begin{bmatrix} 
				\mathbf{C}_{00} & \mathbf{C}_{01} \\
				\mathbf{C}_{10} & \mathbf{C}_{11} 
			\end{bmatrix}.
		\end{align}
		The  matrix $\mathbf{\Xi}$ is a $4 \times 4$ block matrix, where each element $\Xi_{i,j}$ is computed using the Hadamard product (denoted by $\odot$) of specific blocks from matrices $\mathbf{F}$ and $\mathbf{C}$
		\begin{align}
			\Xi_{i,j} = \mathbf{F}_{pq} \odot \mathbf{C}_{mn},
		\end{align}
		where we denote \(\{i_1, i_2\}\) and \(\{j_1, j_2\}\) as the binary representations of \(i-1\) and \(j-1\), respectively. Specifically, the indices \(p\), \(q\), \(m\), and \(n\) are determined as follows: \(p = j_2\), \(q = i_1\), \(m = i_2\), and \(n = j_1\). 
		From the above, we can transform 
		\begin{equation}
			\text{tr}\left(\mathbf{F}\boldsymbol{\Phi}\mathbf{C}\boldsymbol{\Phi}^{\mathrm{H}}\right) = \bm{\phi}^{\mathrm{H}}\boldsymbol{\Xi}\boldsymbol{\phi}.
			\label{eq:46}
		\end{equation}

			For the simplification of $\text{tr}\left(\mathbf{\Phi}\mathbf{P}\right)$ and 
			$ \text{tr}\left(\mathbf{P}^{\mathrm{H}}\boldsymbol{\Phi}^{\mathrm{H}}\right)$,
			 we partition the matrix $\mathbf{P}$ into four blocks
				\begin{align}
			\mathbf{P} = \begin{bmatrix} 
				\mathbf{P}_{00} & \mathbf{P}_{01} \\
				\mathbf{P}_{10} & \mathbf{P}_{11} 
			\end{bmatrix}.
		\end{align}
		Denote $\mathbf{p}_{00}$,$\mathbf{p}_{01}$,$\mathbf{p}_{10}$,$\mathbf{p}_{11}\in\mathbb{C}^{L\times 1}$ as the collection of diagonal elements of matrix $\mathbf{P}_{00}^{\mathrm{H}}$,  $\mathbf{P}_{01}^{\mathrm{H}}$,  $\mathbf{P}_{10}^{\mathrm{H}}$ and  $\mathbf{P}_{11}^{\mathrm{H}}$,  
		and $\bm{\xi} = [\mathbf{p}_{00};\ \mathbf{p}_{10};\ \mathbf{p}_{01};\ \mathbf{p}_{11}]$. Then, we have
		\begin{equation}
			\text{tr}\left(\boldsymbol{\Phi}\mathbf{P}\right) = \bm{\xi}^{\mathrm{H}}\boldsymbol{\phi}.
			\label{eq:47}
		\end{equation}
		Hence,  we can reformulate the optimization problem as
		\begin{subequations}\label{barphi}
			\begin{align}
				&\min_{{\boldsymbol{\phi}}}\ f({\boldsymbol{\phi}}) = {\boldsymbol{\phi}}^{\mathrm{H}}\mathbf{\Xi}{\boldsymbol{\phi}} + 2\Re({\boldsymbol{\phi}}^{\mathrm{H}}\bm{\xi}) \label{barphi_ob}\\
				&\text{s.t.}\quad|{\phi}_j|=1,\quad j=1,\ldots,4L, \label{barphi_cons}
			\end{align}
		\end{subequations}
where the non-convexity of Problem \eqref{barphi} stems from the unit-modulus constraints in \eqref{barphi_cons}. To address this challenge, we employ the MM algorithm as proposed in \cite{Sun2017}. The MM approach consists of two key steps: constructing a surrogate function that approximates the objective function $f({\boldsymbol{\phi}})$, and optimizing it iteratively.
		
\textbf{Lemma 1}: For any feasible ${\boldsymbol{\phi}}$ and current solution ${\boldsymbol{\phi}}^{[t]}$, the following inequality holds
\begin{align}
	\begin{aligned}\label{lemmaphi}
		{\boldsymbol{\phi}}^{\mathrm{H}}\boldsymbol{\Upsilon}_1{\boldsymbol{\phi}} &\leq {\boldsymbol{\phi}}^{\mathrm{H}}\mathbf{X}{\boldsymbol{\phi}} - 2\Re \left\{ {\boldsymbol{\phi}}^{\mathrm{H}}(\mathbf{X} - \mathbf{\Xi}){\boldsymbol{\phi}}^{[t]} \right\} \\
		&\quad + ({\boldsymbol{\phi}}^{[t]})^{\mathrm{H}}(\mathbf{X} - \mathbf{\Xi}){\boldsymbol{\phi}}^{[t]},
	\end{aligned}
\end{align}
		where \( \mathbf{X} = \lambda_{\text{max}}\mathbf{I}_{4L} \) and \( \lambda_{\text{max}} \) is the largest eigenvalue of \( \mathbf{\Xi} \).
After removing constant terms and utilizing Lemma 1, Problem \eqref{barphi_cons} can be rewritten as
		\begin{subequations}\label{asasdcjig}
			\begin{align}
				& \mathop {\max }\limits_{{\boldsymbol{\phi}}} \quad 2\mathop{\rm Re}\left\{ {\boldsymbol{\phi}}^{\rm{H}}\mathbf{q}^{[t]} \right\} \\
				& \textrm{s.t.}\quad |{\phi}_j| = 1 , j = 1, \cdots ,4L, \label{aadshxceur}
			\end{align}
		\end{subequations}
		where \( \mathbf{q}^{[t]} = (\lambda_{\text{max}}\mathbf{I}_{4L} - \mathbf{\Xi}){\boldsymbol{\phi}}^{[t]} -\bm{\xi} \). The optimal solution to Problem \eqref{asasdcjig} is given by
		\begin{equation}\label{dcsd}
			{\boldsymbol{\phi}}^{[t+1]} = e^{j\arg (\mathbf{q}^{[t]})}.
		\end{equation}
This iterative process continues until convergence.
	
		\subsection{Optimization of the DP precoding matrix $\mathbf{W}$}
In the following, we focus on optimizing the DP transmit beamforming matrix $\{\mathbf{F}_k\}$, while keeping $\boldsymbol{\Phi}$, $\left\lbrace {\mathbf{W}}_k\right\rbrace$ and $\left\lbrace {\mathbf{U}}_k\right\rbrace $ fixed. 
By substituting $\mathbf{E}_k$ into \eqref{sum_MSE} and discarding the constant terms, the problem can be reformulated as
\begin{subequations}\label{otp_bk}
	\begin{flalign}
		\mathop {\min }\limits_{\left\lbrace \mathbf{F}_k\right\rbrace }\quad &-\sum_{k=1}^K\operatorname{tr}\left(\mathbf{W}_k\mathbf{U}_k^{\rm{H}}\mathbf{H}_k\mathbf{F}_k\right)\nonumber
		\\
		&+\sum_{k=1}^K\operatorname{tr}\left(\mathbf{F}_k^{\rm{H}}\sum_{m=1}^{K}\mathbf{H}_m^{\rm{H}}\mathbf{U}_m\mathbf{W}_m\mathbf{U}_m^{\rm{H}}\mathbf{H}_m\mathbf{F}_k\right)\nonumber
		\\
		&-\sum_{k=1}^K\operatorname{tr}\left(\mathbf{W}_k\mathbf{F}_k^{\rm{H}}\mathbf{H}_k^{\rm{H}}\mathbf{U}_k\right)
		\label{F_obj}
		\\
		\qquad\  \textrm{s.t.}\quad
		&	(\ref{P1_radar1}),(\ref{P1_radar2}),(\ref{P1_c}).
	\end{flalign}
\end{subequations}

Solving Problem (\ref{otp_bk}) is challenging due to the non-convexity of constraints (\ref{P1_radar1}) and (\ref{P1_radar2}). To address this, a penalty-based method is applied. Specifically, we introduce $\mathbf{X}_k =\mathbf{F}_{k}$, $\mathbf{Y}_k = \mathbf{V}_1\mathbf{F}_{k}$, and $\mathbf{Z}_k = \mathbf{V}_2\mathbf{F}_{k}$. With these auxiliary variables, the problem is reformulated as
\begin{subequations}\label{auxiliary_X_Y}
	\begin{align}
		\mathop {\min }\limits_{\left\lbrace \mathbf{F}_k, \mathbf{X}_k , \mathbf{Y}_k, \mathbf{Z}_k\right\rbrace  } 
		&-\sum_{k=1}^K\operatorname{tr}\left(\mathbf{W}_k\mathbf{U}_k^{\rm{H}}\mathbf{H}_k\mathbf{F}_k\right)\nonumber
		\\
		&+\sum_{k=1}^K\operatorname{tr}\left(\mathbf{F}_k^{\rm{H}}\sum_{m=1}^{K}\mathbf{H}_m^{\rm{H}}\mathbf{U}_m\mathbf{W}_m\mathbf{U}_m^{\rm{H}}\mathbf{H}_m\mathbf{F}_k\right)\nonumber
		\\
		&-\sum_{k=1}^K\operatorname{tr}\left(\mathbf{W}_k\mathbf{F}_k^{\rm{H}}\mathbf{H}_k^{\rm{H}}\mathbf{U}_k\right)
		\label{F_obj20}
		\\
		\textrm{s.t.}\quad
		&\quad\sum_{k=1}^K \operatorname{tr}\left(\mathbf{X}_k\mathbf{X}_k^{\rm{H}}\right) \leq P_0,\label{X_Con}
		\\
		&\quad\sum_{k=1}^{K} \operatorname{tr}\left(\mathbf{Y}_k\mathbf{Y}_k^{\rm{H}}\right)\geq \sigma_\mathrm{r}^{2} \gamma_{\text{1,th}},\label{Y_Con}
		\\
		&\quad\sum_{k=1}^{K} \operatorname{tr}\left(\mathbf{Z}_k\mathbf{Z}_k^{\rm{H}}\right)\geq \sigma_\mathrm{r}^{2} \gamma_{\text{2,th}},\label{Z_Con}\\
		&\quad\mathbf{X}_k = \mathbf{F}_{k} ,\mathbf{Y}_k = \mathbf{V}_1\mathbf{F}_{k},\nonumber\\
		&\quad\mathbf{Z}_k = \mathbf{V}_2\mathbf{F}_{k}.\quad k = 1, \cdots ,K \label{auxiliary}.
	\end{align}
\end{subequations}

The auxiliary constraints in (\ref{auxiliary}) are then incorporated into the objective function as penalty terms, leading to the following penalty-based reformulation
\begin{subequations}\label{Problem_with_auxi_in_obj}
	\begin{align}
		\mathop {\min }\limits_{ {\left\lbrace \mathbf{F}_k  , \mathbf{X}_k ,\mathbf{Y}_k\right\rbrace } } &-\sum_{k=1}^K\operatorname{tr}\left(\mathbf{W}_k\mathbf{U}_k^{\rm{H}}\mathbf{H}_k\mathbf{F}_k\right)\nonumber	-\sum_{k=1}^K\operatorname{tr}\left(\mathbf{W}_k\mathbf{F}_k^{\rm{H}}\mathbf{H}_k^{\rm{H}}\mathbf{U}_k\right)\nonumber
		\\
		&+\sum_{k=1}^K\operatorname{tr}\left(\mathbf{F}_k^{\rm{H}}\sum_{m=1}^{K}\mathbf{H}_m^{\rm{H}}\mathbf{U}_m\mathbf{W}_m\mathbf{U}_m^{\rm{H}}\mathbf{H}_m\mathbf{F}_k\right)\nonumber
		\\	
		&+\frac{1}{2\rho^{\left[ t\right] }}\bigg( \sum_{k=1}^K \Vert {\mathbf{X}_k} - \mathbf{F}_{k}\Vert_{F}^{2}+ \sum_{k=1}^K \Vert {\mathbf{Y}_k} -  \mathbf{V}_1\mathbf{F}_{k}\Vert_{F}^{2}\nonumber
		\\
		&+\sum_{k=1}^K \Vert {\mathbf{Z}_k} -  \mathbf{V}_2\mathbf{F}_{k}\Vert_{F}^{2}\bigg) \label{obj_value}
		\\
		\textrm{s.t.}\quad\
		&(\ref{X_Con}),(\ref{Y_Con}),(\ref{Z_Con}).
	\end{align}
\end{subequations}
Here, $\rho^{\left[ t\right]} \ (\rho^{\left[ t\right]} > 0)$ represents the penalty parameter at the $t$-th iteration. During the outer-layer process, this penalty parameter is gradually reduced to encourage the solution to closely satisfy the auxiliary constraints (\ref{auxiliary}) while minimizing the objective function. By driving $\rho^{\left[ t\right]}$ towards zero in the outer iterations, the penalty coefficient $\frac{1}{2 \rho^{\left[ t\right] }}$ increases towards infinity. This ensures that the auxiliary constraints (\ref{auxiliary}) are satisfied with increasing precision. This approach balances the dual goals of minimizing the original objective and enforcing the equality constraints.

In the inner layer of the penalty-based algorithm, the penalty coefficient $\rho^{\left[ t\right]}$ remains fixed. To handle the non-convexity, we split the problem of the inner layer into two subproblems: one optimizes the beamforming matrix, and the other focuses on the auxiliary matrices. These subproblems are alternately solved until convergence.

		\subsubsection{Outer Layer Update}
		 In the outer loop, the penalty coefficient is updated according to
		 \begin{align}
		 	\rho^{[m]} = c\rho^{[m-1]},
		 \end{align}
		  where $0 < c < 1$ serves as a step-size factor. 
		   Without loss of generality, we will now focus on the 
		  $m$-th outer layer update as an example to specifically introduce the inner layer optimization process.
		
		\subsubsection{Inner Layer Optimization}
		\paragraph{With fixed $\{\mathbf{X}_k, \mathbf{Y}_k, \mathbf{Z}_k\}$, Optimize $\{\mathbf{F}_k\}$} the problem simplifies to an unconstrained convex quadratic problem
		\begin{align}\label{admmobj}
			\begin{aligned}
				\min_{\{\mathbf{F}_k\}}\ 
				&-\sum_{k=1}^K\operatorname{tr}\left(\mathbf{W}_k\mathbf{U}_k^{\rm{H}}\mathbf{H}_k\mathbf{F}_k\right)	-\sum_{k=1}^K\operatorname{tr}\left(\mathbf{W}_k\mathbf{F}_k^{\rm{H}}\mathbf{H}_k^{\rm{H}}\mathbf{U}_k\right)
				\\
				&+\sum_{k=1}^K\operatorname{tr}\left(\mathbf{F}_k^{\rm{H}}\sum_{m=1}^{K}\mathbf{H}_m^{\rm{H}}\mathbf{U}_m\mathbf{W}_m\mathbf{U}_m^{\rm{H}}\mathbf{H}_m\mathbf{F}_k\right)
				\\	
				&+\frac{1}{2\rho^{\left[ t\right] }}\bigg( \sum_{k=1}^K \Vert {\mathbf{X}_k} - \mathbf{F}_{k}\Vert_{F}^{2}+ \sum_{k=1}^K \Vert {\mathbf{Y}_k} -  \mathbf{V}_1\mathbf{F}_{k}\Vert_{F}^{2}
				\\
				&+\sum_{k=1}^K \Vert {\mathbf{Z}_k} -  \mathbf{V}_2\mathbf{F}_{k}\Vert_{F}^{2}\bigg).
			\end{aligned}
		\end{align}
		The optimal solution can be obtained through the first-order optimality conditions. Let
		\begin{align}
			\mathbf{A}_k = &\bigg(  2\sum_{m=1}^{K}\mathbf{H}_m^{\rm{H}}\mathbf{U}_m\mathbf{W}_m\mathbf{U}_m^{\rm{H}}\mathbf{H}_m  +\\\nonumber
			 &\frac{1}{\rho^{(t)}} \left( \mathbf{I}_{N_{\rm{ t}}}  + \mathbf{V}_1^{\rm{H}}\mathbf{V}_1 + \mathbf{V}_2^{\rm{H}}\mathbf{V}_2\right)  \bigg) \\
			\mathbf{b}_k = &\bigg(\frac{1}{\rho^{(t)}}\left( \mathbf{X}_{k} + \mathbf{V}_1^{\rm{H}}\mathbf{Y}_k+\mathbf{V}_2^{\rm{H}}\mathbf{Z}_k\right) + 2\mathbf{H}_{k}^{\rm{H}}\mathbf{U}_{k}\mathbf{W}_{k} \bigg),\label{bk}\
		\end{align}
		thus, the optimal solution can be derived as
		\begin{align}\label{calF}
			\mathbf{F}_k = \mathbf{A}_k^{-1} \mathbf{b}_k\quad k=1,\cdots,K.
		\end{align}

		\paragraph{Optimizing $\left\{\mathbf{X}_k\right\}$ with Fixed $\{\mathbf{F}_k,\mathbf{Y}_k,\mathbf{Z}_k\}$}
		
		\begin{algorithm}
			\caption{Bisection Search Method with Complementary Slackness}
			\label{alg-modified}
			\begin{algorithmic}[1]
				\State \textbf{Initialize} the accuracy $\eta$, the bounds $\tau_{\text{lb}} = 0$ and $\tau_{\text{ub}} = \sqrt{\frac{\Lambda}{P_0}}$.
				\If {$g(0) \leq P_0$} 
				\State Set $\tau^{\text{opt}} = 0$, and $\mathbf{X}_k^{\text{opt}} = \mathbf{F}_k$ for all $k$. Terminate.
				\EndIf
				\While {$\tau_{\text{ub}} - \tau_{\text{lb}} > \eta$}
				\State Compute $\tau = \frac{\tau_{\text{lb}} + \tau_{\text{ub}}}{2}$.
				\State Evaluate $g(\tau) = \frac{\Lambda}{(1+\tau)^2}$.
				\If {$g(\tau) > P_0$}
				\State Set $\tau_{\text{lb}} = \tau$.
				\Else
				\State Set $\tau_{\text{ub}} = \tau$.
				\EndIf
				\EndWhile
				\State \textbf{Output} $\tau^{\text{opt}} = \frac{\tau_{\text{lb}} + \tau_{\text{ub}}}{2}$
			\end{algorithmic}
		\end{algorithm}
		The subproblem with $\{\mathbf{X}_k\}$ becomes
		\begin{subequations}\label{calX}
			\begin{align}
				\mathop {\min }\limits_ {\left\lbrace \mathbf{X}_k\right\rbrace }\quad 
				&\sum_{k=1}^K\Vert {\mathbf{X}_k} - \mathbf{F}_{k}\Vert_{F}^{2}
				\\
				\quad\textrm{s.t.}\quad
				&(\ref{X_Con}).\label{ADMM_X_c}
			\end{align}
		\end{subequations}
		This problem is convex and can be addressed by solving its dual problem. The Lagrangian function with Lagrange multiplier $\tau$ ($\tau \geq 0$) is
		\begin{equation}\label{L1}
			\mathcal{L}_1\left(\mathbf{X}_{k}, \tau \right)=\sum_{k=1}^{K}\Vert {\mathbf{X}_k} - \mathbf{F}_{k}\Vert^{2}_{F} +\tau\left(\sum_{k=1}^K \operatorname{tr}\left(\mathbf{X}_k\mathbf{X}_k^{\rm{H}}\right) - P_0\right).
		\end{equation}
		According to Karush-Kuhn-Tucker (KKT) conditions, Its dual function, $f_1(\tau) = \min_{\{\mathbf{X}_k\}} \mathcal{L}_1(\{\mathbf{X}_k\}, \tau)$, is bounded if $\tau \ge 0$. The gradient of the Lagrangian $\mathcal{L}_1$ with respect to $\mathbf{X}_k$ is set to 0, 
		which leads to the closed-form solution
		\begin{align}\label{X_k_solutions}
			\mathbf{X}_k^{\text{opt}}(\tau) = \frac{\mathbf{F}_k}{1+\tau}, \quad k = 1, \ldots, K,
		\end{align}
		where $\tau$ is determined by
		\begin{align}\label{X_g}
			g(\tau) & \triangleq \sum_{k=1}^{K} \operatorname{tr}(\mathbf{X}_k(\tau)\mathbf{X}_k^{\mathrm{H}}(\tau)) \nonumber 
			= \frac{\Lambda}{(1+\tau)^{2}},
		\end{align}
		with $\Lambda = \sum_{k=1}^{K}\text{tr}(\mathbf{F}_k^{\mathrm{H}}\mathbf{F}_k)$. The function $g(\tau)$ is monotonically decreasing for $\tau \geq 0$.  According to the complementary slackness condition, the product of the dual variable $\tau$ and the constraint violation must be zero, i.e.,
		\begin{align}
			\tau \cdot \left(\sum_{k=1}^K \operatorname{tr}(\mathbf{X}_k \mathbf{X}_k^{\mathrm{H}}) - P_0\right) = 0.
		\end{align}
		If $g(0) \leq P_0$, then $\tau^{\text{opt}} = 0$. Otherwise, it can be found by solving
	\begin{equation}
		g(\tau) = \frac{\Lambda}{(1+\tau)^{2}} = P_0.
	\end{equation}
	The upper bound for \(\tau\), denoted by \(\tau_{\text{ub}}\), should satisfy the following condition
	\begin{equation}\label{prove_tau_up}
		g(\tau_{\text{ub}}) = \frac{\Lambda}{(1+\tau_{\text{ub}})^2} < \frac{\Lambda}{\tau_{\text{ub}}^2} = P_0.
	\end{equation}
	From this, we can derive the upper bound as \(\tau_{\text{ub}} = \sqrt{\frac{\Lambda}{P_0}}\).
	
	By exploiting the monotonic property of $g(\tau)$, the solution $\tau^{\text{opt}}$ can be easily obtained using a straightforward bisection search method in Algorithm \ref{alg-modified}.

		\begin{algorithm}
			\caption{Penalty-based algorithm for solving Problem (\ref{otp_bk}).}\label{Alg-pdd}
			\begin{algorithmic}[1]
				\State {Initialize $\mathbf{F}^{\left[ 0 \right] } =  \mathbf{F}^{\left( \mathrm{n}\right) }$ ,$  \mathbf{X}^{\left[  0\right]  }  =  \mathbf{F}^{\left( \mathrm{n}\right) }$, $
					 \mathbf{Y}^{\left[ 0\right]  } =  \mathbf{V}_1\mathbf{F}_{k}^{\left( \mathrm{n}\right) }$, $
					 \mathbf{Z}^{\left[ 0\right]  } =  \mathbf{V}_2\mathbf{F}_{k}^{\left( \mathrm{n}\right) }$, $m=0$, the penalty coefficient $\rho^{[0]}$, step size $c$ , tolerance of accuracy $\xi$ and $\epsilon$.}
				\While{the fractional increase of (\ref{admmobj}) is larger than $\epsilon$ or penalty terms larger than $\xi$}
				\State {Update  $\mathbf{F}^{[m+1]}$ by using (\ref{calF});}
				\State {Update   $\mathbf{X}^{[m+1]}$ by solving Problem (\ref{calX}); }
				\State {Update  $\mathbf{Y}^{[m+1]}$ by solving Problem (\ref{calY});}
				\State {Update  $\mathbf{Z}^{[m+1]}$ by solving Problem (\ref{calZ});}
				\State {Set $\rho^{[m+1]}=c \rho^{[m]}, m \leftarrow m+1 $.}
				\EndWhile
				\State \textbf{Output} $ \mathbf{F}^{ \left( n+1 \right)  }  =  \mathbf{F}^{ \left[ m\right]  } $.
			\end{algorithmic}
		\end{algorithm}

				\paragraph{Optimizing $\left\{\mathbf{Y}_k\right\}$ with Fixed $\{\mathbf{F}_k,\mathbf{X}_k,\mathbf{Z}_k\}$}
The subproblem associated with $\left\{\mathbf{Y}_k\right\}$ is given by
\begin{subequations}\label{calY}
	\begin{align}
		\min_{\left\{\mathbf{Y}_k\right\}}\quad 
		&\sum_{k=1}^K\Vert \mathbf{Y}_k - \mathbf{V}_1\mathbf{F}_{k}\Vert_{F}^{2} \\
		\textrm{s.t.}\quad
		&(\ref{Y_Con}). \label{ADMM_Y_c}
	\end{align}
\end{subequations}
This problem is a Quadratic Constraint Quadratic Programming (QCQP) problem. Specifically, we formulate the Lagrangian function with multiplier $\mu_1 \geq 0$ to solve this QCQP problem
\begin{align}\label{L_2}
	\mathcal{L}_2(\mathbf{Y}_k, \mu_1) = & (1 - \mu_1) \sum_{k=1}^K \Vert \mathbf{Y}_k \Vert_{F}^{2} - \sum_{k=1}^K 2\operatorname{Re}(\mathbf{F}_k^{\mathrm{H}}\mathbf{V}_1^{\mathrm{H}}\mathbf{Y}_k) \nonumber\\
	& + \sum_{k=1}^K \Vert \mathbf{V}_1\mathbf{F}_{k} \Vert_{F}^{2} + \mu_1\sigma_{r}^2\gamma_{\text{1,th}}.
\end{align}
Define $f_2(\mu_1) = \min_{\{\mathbf{Y}_k\}} \mathcal{L}_2(\{\mathbf{v}_k\}, \mu_1)$ as the dual function of Problem \eqref{L_2}.
To ensure the boundedness of $f_2(\mu_1)$, we must restrict $\mu_1$ to the range $0 \leq \mu_1 < 1$. 
If \( \mu_1 > 1 \), the term \( (1 - \mu_1) \) in the Lagrangian becomes negative. In this case, setting \( \mathbf{Y}_k = \rho \mathbf{1}_{N_t} \) and letting \( \rho \to \infty \) causes the dual function \( f_2(\mu_1) \) to drop endlessly, making it unbounded.
If \( \mu_1 = 1 \), the term \( (1 - \mu_1) \) becomes zero, which creates undefined or unstable behavior.

Using the first-order optimality conditions, the optimal \( \mathbf{Y}_k \) is given by
\begin{eqnarray}\label{u_k_solution}
	\mathbf{Y}_k^{\text{opt}}(\mu_1) = \frac{\mathbf{V}_1\mathbf{F}_k}{1-\mu_1}, \quad k=1,\ldots,K.
\end{eqnarray}
To determine the optimal $\mu_1$, we define the auxiliary function
\begin{eqnarray}\label{h_u}
	h(\mu_1,\{\mathbf{Y}_k\},\mathbf{V}_1) &\triangleq& \sum_{k=1}^{K} \operatorname{tr}(\mathbf{Y}_k(\mu_1)\mathbf{Y}_k^{\mathrm{H}}(\mu_1)) \nonumber \\
	&=& \operatorname{tr}\big((1-\mu_1)^{-2} \mathbf{\Delta}(\{\mathbf{Y}_k\},\mathbf{V}_1)\big) \nonumber \\
	&=& \sum_{i=1}^{N_{\text{t}}}\frac{[\mathbf{\Delta}(\{\mathbf{Y}_k\},\mathbf{V}_1)]_{i,i}}{(1-\mu_1)^{2}},
\end{eqnarray}
where $\mathbf{\Delta}(\{\mathbf{Y}_k\},\mathbf{V}_1) = \sum_{k=1}^{K}\mathbf{V}_1\mathbf{Y}_k\mathbf{Y}_k^{\mathrm{H}}\mathbf{V}_1^{\mathrm{H}}$ and $[\mathbf{\Delta}(\{\mathbf{Y}_k\},\mathbf{V}_1)]_{i,i}$ denotes the $i$-th diagonal element. The function $h(\mu_1)$ is strictly increasing for $\mu_1 \in [0,1)$. The optimal value of $\mu_1$ is determined as follows
\begin{equation}\label{eq:mu-optimal}
	\mu_1^{\text{opt}} = \begin{cases}
		0, & \text{if}\ h(0,\{\mathbf{Y}_k\},\mathbf{V}_1) \geq \gamma_{\text{1,th}}\sigma_r^2 \\
		\mu_1^*, & \text{otherwise}
	\end{cases},
\end{equation}
where $\mu_1^*$ is the solution to $h(\mu_1,\{\mathbf{Y}_k\},\mathbf{V}_1) = \gamma_{\text{1,th}}\sigma_r^2$, which can be efficiently found using the bisection search in the interval $[ 0 , 1 )$.
		\paragraph{Optimizing $\left\{\mathbf{Z}_k\right\}$ with Fixed $\{\mathbf{F}_k,\mathbf{X}_k,\mathbf{Y
			}_k\}$} The subproblem associated with $\left\{\mathbf{Z}_k\right\}$ is 
		\begin{subequations}\label{calZ}
			\begin{align}
				\min_{\left\{\mathbf{Z}_k\right\}}\quad 
				&\sum_{k=1}^K\Vert \mathbf{Z}_k - \mathbf{V}_2\mathbf{F}_{k}\Vert_{F}^{2} \\
				\textrm{s.t.}\quad
				&(\ref{Z_Con}). \label{ADMM_Z_c}
			\end{align}
		\end{subequations}
		This problem is similar to Problem \eqref{calY}, using the same method, we have
	\begin{eqnarray}
		\mathbf{Z}_k^{\text{opt}}(\mu_2) = \frac{\mathbf{V}_2\mathbf{F}_k}{1-\mu_2}, \quad k=1,\ldots,K.
	\end{eqnarray}
	 The optimal value of $\mu_2$ is
	\begin{equation}
		\mu_2^{\text{opt}} = \begin{cases}
			0, & \text{if}\ h(0,\{\mathbf{Z}_k\},\mathbf{V}_2) \geq \gamma_{\text{2,th}}\sigma_r^2 \\
			\mu_2^*, & \text{otherwise}
		\end{cases},
	\end{equation}
	where $\mu_2^*$ is the solution to $h(\mu_2) = \gamma_{\text{2,th}}\sigma_r^2$.

		\begin{algorithm}
			\caption{Overall Algorithm for DP RIS-Aided ISAC Systems}
			\label{overallAlg}
			\begin{algorithmic}[1]
				\State Initialize $\mathbf{F}^{(1)}$, $\mathbf{\Phi}^{(1)}$, and set iteration index $n=1$.
				\State Set convergence threshold $\epsilon$ and maximum number of iterations $N_{\text{max}}$.
				\State Given  $\mathbf{F}^{(n)}$ and $\mathbf{\Phi}^{(n)}$, calculate $\mathbf{U}^{(n)}$ in (\ref{eq:17}); 
				
				\State  Given  $\mathbf{F}^{(n)}$, $\mathbf{\Phi}^{(n)}$ and $\mathbf{U}^{(n)}$, calculate $\mathbf{W}^{(n)}$ in (\ref{eq:18}); 
				\State Given  $\mathbf{F}^{(n)}$, $\mathbf{\Phi}^{(n)}$, $\mathbf{U}^{(n)}$ and $\mathbf{W}^{(n)}$,
				update $\mathbf{\Phi}^{(n+1)}$ in (\ref{dcsd});
				\State Given  $\mathbf{U}^{(n)}$, $\mathbf{W}^{(n)}$ and $\mathbf{\Phi}^{(n)}$, update $\mathbf{F}^{(n+1)}$ through Algorithm \ref{Alg-pdd};
				\State Calculate $\delta=\max(\|\mathbf{\Phi}^{(n+1)} - \mathbf{\Phi}^{(n)}\|, \|\mathbf{F}^{(n+1)} - \mathbf{F}^{(n)}\|)$;
				\State If {$\delta < \epsilon$ \textbf{or} $n \geq N_{\text{max}}$}, terminate. Otherwise, set $n \gets n + 1$ and go to step 3.

			\end{algorithmic}
		\end{algorithm}

		\subsection{Complexity analysis}
		The overall algorithm is summarized in Algorithm \ref{overallAlg}. In the following, we analyze the complexity of our proposed methods, the main computation load lies in optimizing two variables $\bm{\Phi}$ and $\mathbf{W}$. 
	In step 3, the complexity of computing the decoding matrices $\mathbf{U}^{(n)}$ is given by  $o_1=\mathcal{O}(8K)$. In step 4,  the complexity of computing the auxiliary matrices $o_2=\mathbf{W}^{(n)}$ is  $\mathcal{O}(8K)$.
	In step 5, we need to calculate the maximum eigenvalue of $\Xi_1$, whose associated complexity is $\mathcal{O}(8L^3)$. For each iteration of the MM algorithm, the calculation of $\mathbf{q}^{[t]}$ needs the complexity of $\mathcal{O}(4L^2)$. The time for the convergence of the MM algorithm is $T_{\text{MM}}$. The complexity for $\bm{\Phi}$ is $o_3=\mathcal{O}(8L^3+4T_{\text{MM}}L^2)$. In step 6, the maximum iteration number of the out and inner layers are set to be $I_{\text{max}}^o$ and $I_{\text{max}}^{i}$. Main computing resources are calculating $\mathbf{W}$, $\Lambda$ and $\mathbf{\Delta}$, whose complexity is $\mathcal{O}(KN_t^3)$,$\mathcal{O}(KN_t)$ and $\mathcal{O}(KN_t^2)$. Ignoring the computation complexity in the outer layer, the complexity of step 6 is $o_4= \mathcal{O}(I_{\text{max}}^oI_{\text{max}}^i(KN_t^3+KN_t+KN_t^2))$. 
	In summary, the overall complexity of the proposed algorithm is $o_1+o_2+o_3+o_4$.

		\section{SIMULATION RESULTS}
		In this section, we evaluate the performance of the DP RIS-aided ISAC systems through extensive simulation results. The large-scale path loss in dB follows the standard model
		\begin{align}
			\text{PL}=\text{PL}_0-10\alpha\log{10}(\frac{d}{d_0}),
		\end{align}
		where $\text{PL}_0 = -30$ dB represents the path loss at the reference distance $d_0=1$ m. The path loss exponents are set to $\alpha_{\text{BS-RIS}} = 2.25$, $\alpha_{\text{RIS-UE}} = 2.25$, and $\alpha_{\text{BS-UE}} = 4.75$, respectively.

		For the direct channel from the DP BS to UEs, the small-scale fading is assumed to be Rayleigh fading due to extensive scatters. Each element of sub-channel $\mathbf{h}_{d,k,pq}$ and  $\mathbf{h}_{d,k,pp}$ follows the distribution of $\mathcal{CN}(0,\beta_1)$ and $\mathcal{CN}(0,1-\beta_1)$, respectively.
		For the RIS-related channels (BS-RIS and RIS-UEs),  we assume that they follow Rician fading which are given by \eqref{Gchannel} and \eqref{hchannel}. Specifically, the shared NLoS channel follows Rayleigh fading and the shared LoS channel in \eqref{Gchannel} and \eqref{hchannel} is modeled as
		\begin{equation}
			\hat{\mathbf{H}}^{\text{LoS}} =\mathbf{a}_{r}(\vartheta^{\text{AoA}})\mathbf{a}_{t}^{\mathrm{H}}(\vartheta^{\text{AoD}}),
			\label{eq:channel_model}
		\end{equation}
		where the array response vectors $\mathbf{a}_{r}(\vartheta^{\text{AoA}})$ and $\mathbf{a}_{t}(\vartheta^{\text{AoD}})$ are defined as that in \eqref{steervector}
		, and $\vartheta^{\text{AoA}}$ and $\vartheta^{\text{AoD}}$ denote the angle of arrival and angle of departure, respectively. 
		The XPD for LoS and NLoS are set to 8 dB and 5 dB, so we have $\beta_2^{\text{LoS}}=\beta_3^{\text{LoS}}=0.1368$ and $\beta_1^{\text{NLoS}}=\beta_2^{\text{NLoS}}=\beta_3^{\text{NLoS}}=0.2403$.
		
\begin{figure}[bhtp]
			\centering\vspace*{-0.50\baselineskip}
			\includegraphics[width=3.9in]{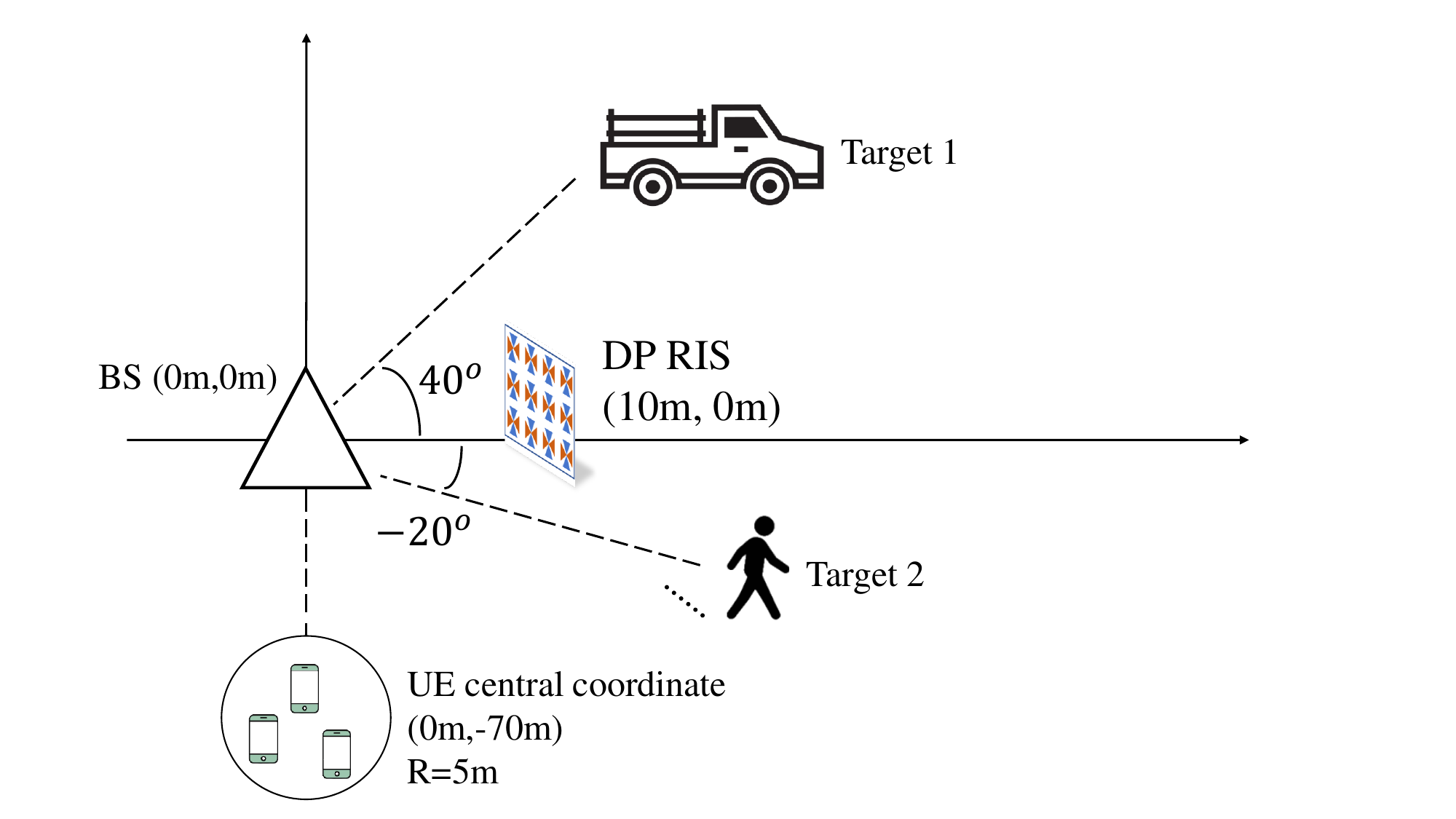}
			\caption{The simulation system scenario.}
			\label{isacppt}
		\end{figure}

		As illustrated in Fig. \ref{isacppt}, the 2D coordinates of the BS and the RIS are $(0\ \text{m}, 0\ \text{m})$ and $(10\ \text{m}, 0\ \text{m})$, respectively. The center coordinates of the UEs are $(0\ \text{m}, -70\ \text{m})$, the user distribution radius is $5\ \text{m}$. The targets 1 and 2 are located at $-20^{\circ}$ and $40^{\circ}$ away from the BS, the distance is 100 m. The simulation system scenario is displayed in Fig. \ref{isacppt}.

Unless otherwise specified, the following simulation parameters are adopted: Channel bandwidth of $10$ MHz, noise power density of $-174$ dBm/Hz, number of DP transmit antennas of $N_t=6$, number of receive antennas for sensing of $N_r=6$,  number of DP relection elements of $L=10$, number of users of $K=3$, maximum DP BS power of $P_{0}=1$, Rician factor is 2.5. When optimizing $\mathbf{F}$, the initial value of $\rho$ is set to 5, and the step size $c$ is set to 0.7.

		\subsection{Convergence of the proposed algorithm}
			\begin{figure}[t]
			\centering\vspace*{-0.50\baselineskip}
			\includegraphics[width=3.5in]{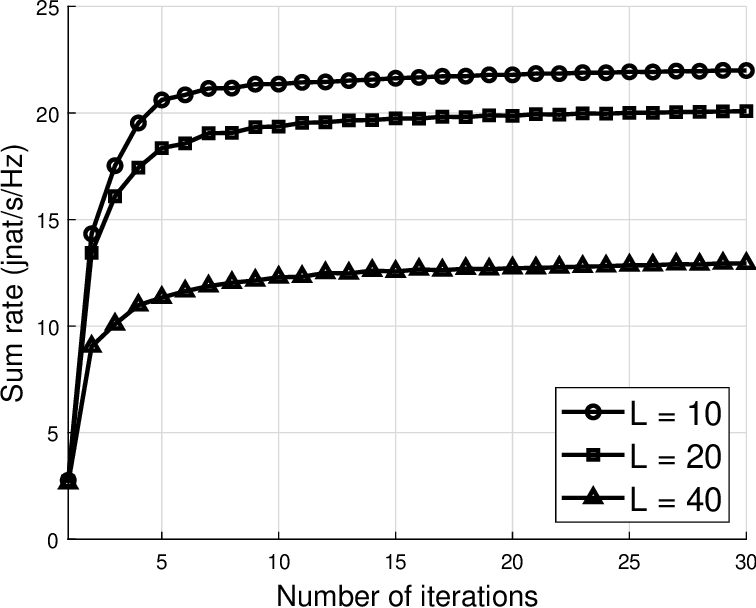}
			\caption{Convergence behavior of the proposed algorithm}
			\label{WSRvsRISnum}
		\end{figure}
		
		\begin{figure}[t]
			\centering\vspace*{-0.50\baselineskip}
			\includegraphics[width=3.5in]{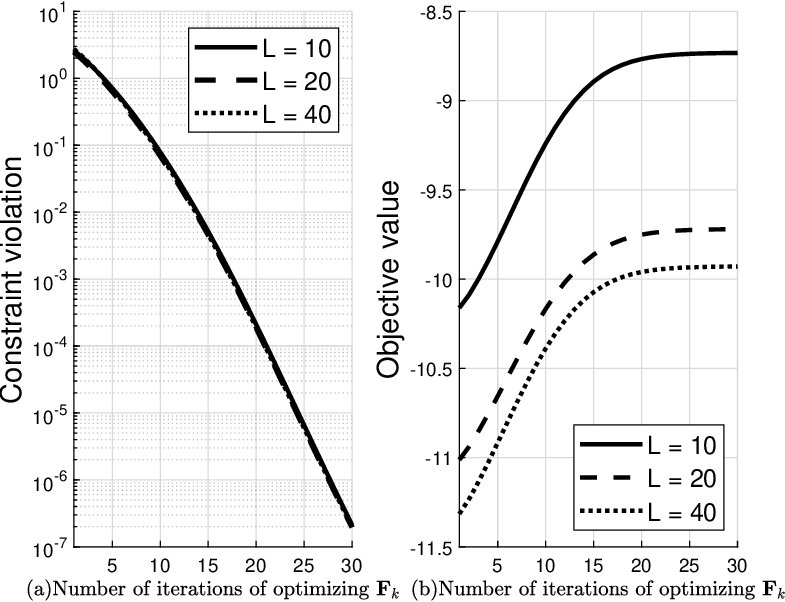}
			\caption{Convergence behavior of the constraints in Algorithm \ref{Alg-pdd}}
			\label{ConstrWSRvsRISnum}
		\end{figure}
	
	We present the convergence behavior of the proposed algorithm in Fig. \ref{WSRvsRISnum}, where the number of DP RIS array elements is set to 10, 20, and 40, respectively. The results reveal that the algorithm exhibits a relatively fast convergence behavior, especially after the first ten iterations.  Moreover, it is evident that the sum rate correspondingly increases as the number
	of DP RIS elements increases. 
	
	Also, we evaluate the convergence behavior of Algorithm \ref{Alg-pdd} with two key components. The objective function, as shown in Fig. \ref{ConstrWSRvsRISnum}, achieves convergence during approximately 15 iterations. As the penalty parameter $\rho$ decreases iteratively, a corresponding reduces in the penalty terms $\sum_{k=1}^{K} \|\mathbf{X}_k - \mathbf{V}_1 \mathbf{F}_k \|_F^2$, $\sum_{k=1}^{K} \|\mathbf{Y}_k - \mathbf{V}_2 \mathbf{F}_k \|_F^2$, and $\sum_{k=1}^{K} \|\mathbf{Z}_k - \mathbf{F}_k \|_F^2$. These terms ultimately converge to approximately $10^{-6}$ after 30 iterations. This negligible residual value demonstrates that the equality constraint is satisfied with high precision.

	\subsection{Comparison with SP scenarios}
			\begin{figure}[t]
		\centering\vspace*{-0.50\baselineskip}
		\includegraphics[width=3.5in]{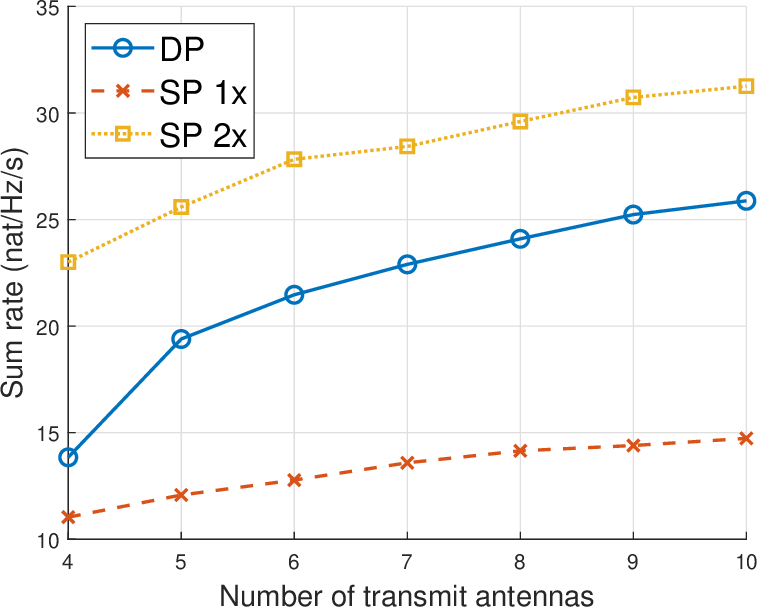}
		\caption{Comparison between SP and DP scenarios}
		\label{WSRvsuni}
	\end{figure}
	In order to obtain more insights about the benefits of considering DP setup, we conduct two comparative experiments with DP scenarios in Fig. \ref{WSRvsuni}. Following the experimental setup in \cite{ozdogan2022massive}, we consider the following configurations:
	
\textbf{1)} SP 1x: In this scenario, both transmit antennas and RIS elements adopt SP configurations, with the number of antennas and RIS elements equal to those in the DP scenario, and each user is equipped with an SP antenna. This configuration directly demonstrates the performance advantages of DP systems under the same hardware scale.

\textbf{2)} SP 2x: To ensure fair comparison regarding the total system degrees of freedom, this configuration employs twice the number of DP transmit antennas and DP RIS elements, and each user is equipped with two SP antennas.

For SP channels, the polarization leakage impact in the SP MIMO system differs from the DP case when assuming identical polarization across all antennas. $\bar{\mathbf{H}}_{d,k}$, $\bar{\mathbf{G}}$ and $\bar{\mathbf{H}}_{r,k}$ denote the BS-UEs link, BS-RIS link, RIS-UEs link for SP scenarios, respectively. Consequently, the elements of $\bar{\mathbf{H}}_{d,k}$ follow the distribution of $\mathcal{CN}(0,1-\beta_1)$, and
\begin{align}
	&\bar{\mathbf{G}} = \omega_{\text{L}}\sqrt{1-\beta_2^{\text{LoS}}}\bar{\mathbf{G}}^{\text{LoS}}+\omega_{\text{N}}\sqrt{1-\beta_2^{\text{NLoS}}}\bar{\mathbf{G}}^{\text{NLoS}},\\
	&\bar{\mathbf{H}}_{r,k} = \omega_{\text{L}}\sqrt{1-\beta_3^{\text{LoS}}}\bar{\mathbf{H}}_{r,k}^{\text{LoS}}+\omega_{\text{N}}\sqrt{1-\beta_3^{\text{NLoS}}}\bar{\mathbf{H}}_{r,k}^{\text{NLoS}},\
\end{align}
where $\bar{\mathbf{G}}^{\text{LoS}}$, $\bar{\mathbf{G}}^{\text{NLoS}}$, $\bar{\mathbf{H}}_{r,k}^{\text{LoS}}$ and $\bar{\mathbf{H}}_{r,k}^{\text{NLoS}}$ are defined similarly to those in DP scenarios.

The simulation results demonstrate the comparison of DP, SP 1x and SP 2x scenarios across various transmit antenna scales (4-10 elements) and the number of DP RIS elements is 10.
DP implementation (blue) achieves superior spectral efficiency, scaling from 14 to 26 nat/Hz/s as $N_t$ increases from 4 to 10. While SP 2x (yellow) reaches higher absolute throughput (23-31 nat/Hz/s), it requires twice the physical dimensions compared with our proposed DP system. SP 1x (red) shows limited scaling (11-15 nat/Hz/s).
This result is similar to the results in \cite{ozdogan2022massive}. It can be seen that 
in the same physical size, the communication rate of the DP configuration is twice that of the SP configuration, and it increases rapidly as the number of transmitting antennas increases. Finally, due to the depolarization effect and antenna aperture, the performance of DP system is slightly worse than that of an SP system with twice the physical size.

	\subsection{Impact of quantization bits on system performance}
	\begin{figure}[t]
		\centering\vspace*{-0.50\baselineskip}
		\includegraphics[width=3.5in]{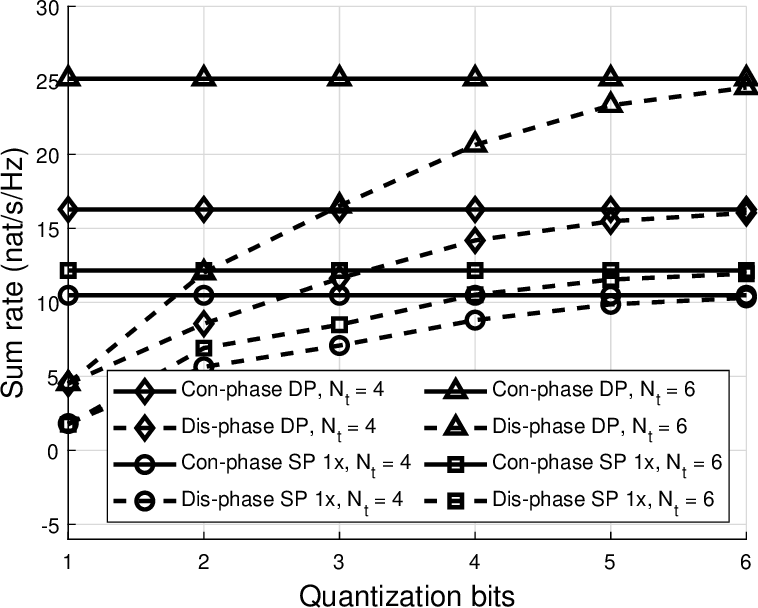}
		\caption{Sum rate versus the quantization Bits of DP RIS}
		\label{WSRquan}
	\end{figure}
	In practical RIS-aided communication systems, continuous phase (`con-phase') cannot be achieved. Therefore, we investigate the impact of discrete phase  (`dis-phase') shift quantization on the system sum rate performance. The discrete phase shifts are obtained by quantizing the optimal continuous phase shifts using $M$ bits, where the discrete phase-shift set is defined as
\begin{equation}
	\mathcal{A} = \left\{\frac{2\pi m}{2^M}\right\}, m \in [0, 2^M - 1], m \in \mathbb{N}.
\end{equation}
For a given optimal continuous phase shift $\phi^*(n)$ obtained by our proposed algorithm, the corresponding discrete phase shift $\hat{\phi}(n)$ is determined by 
$
\hat{\phi}(n) = \underset{\phi \in \mathcal{A}}{\text{argmin}} \{|\phi - \phi^*(n)|\}.
$

Fig.~\ref{WSRquan} presents the system sum rate performance versus the number of quantization bits $M$ under different numbers of transmit antennas $N_t$. $N_t$ is set to 6 and 8, and the number of DP RIS elements is 40. Also, we set the configuration ``SP 1x'' scenario for comparison.    From the simulation results, we observe that the sum rate rapidly increases with $M$ when the number of quantization bits is small ($M \leq 3$), indicating that increasing the quantization precision initially brings significant performance improvements. As $M$ increases further ($M > 4$), the performance gain becomes marginal, and the sum rate gradually approaches that of the continuous phase shifts. This suggests a diminishing return in performance improvement with increased quantization precision.
Moreover, the system with larger $N_t$ consistently achieves higher sum rates and the DP system has greater rate growth potential. By comparison, we show that the DP system still maintains good performance in the case of phase quantization.

\subsection{Impact of  XPD of DP channel on system performance}
\begin{figure}[t]
	\centering\vspace*{-0.50\baselineskip}
	\includegraphics[width=3.5in]{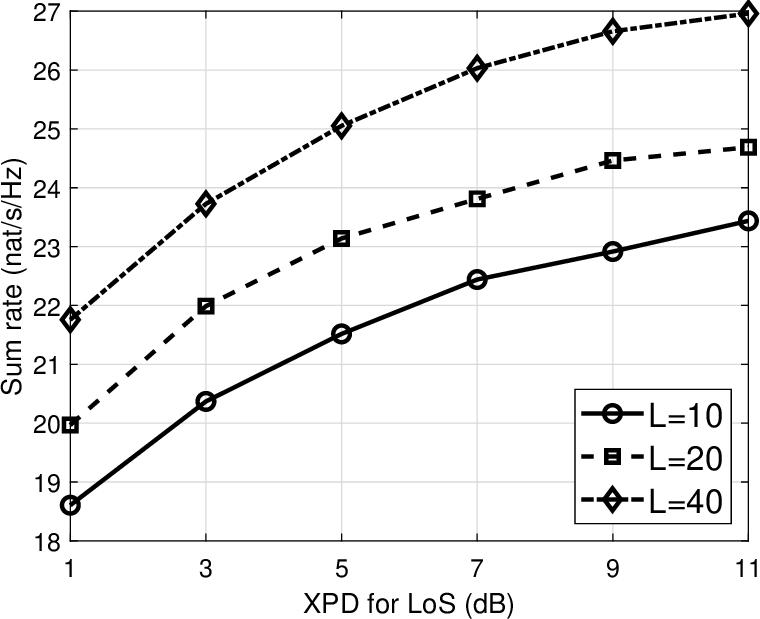}
	\caption{Sum rate versus the XPD of DP channel}
	\label{WSRXPD}
\end{figure}
Fig. \ref{WSRXPD} demonstrates the  impact of XPD of the DP channel on the system's sum rate performance. The simulation is configured with XPD for LoS as $[1, 3, 5, 7, 9, 11]$ dB and the corresponding XPD for NLoS as $[4,6,8,11,13,14]$ dB. The results reveal that higher XPD values lead to substantial improvements in sum rate across all RIS configurations: with $L=40$, the sum rate improves from 21.7 to 27 nat/s/Hz; the $L=20$ configuration shows an improvement from 20 to 24.8 nat/s/Hz; while $L=10$ achieves an enhancement from 18.6 to 23.5 nat/s/Hz. Additionally, larger $L$ values demonstrate better absolute performance and higher sensitivity to XPD improvements, as shown by the steeper slope of the $L=40$ curve.

\subsection{Impact of the sensing SNR on the sum rate}

\begin{figure}[t]
	\centering\vspace*{-0.50\baselineskip}
	\includegraphics[width=3.5in]{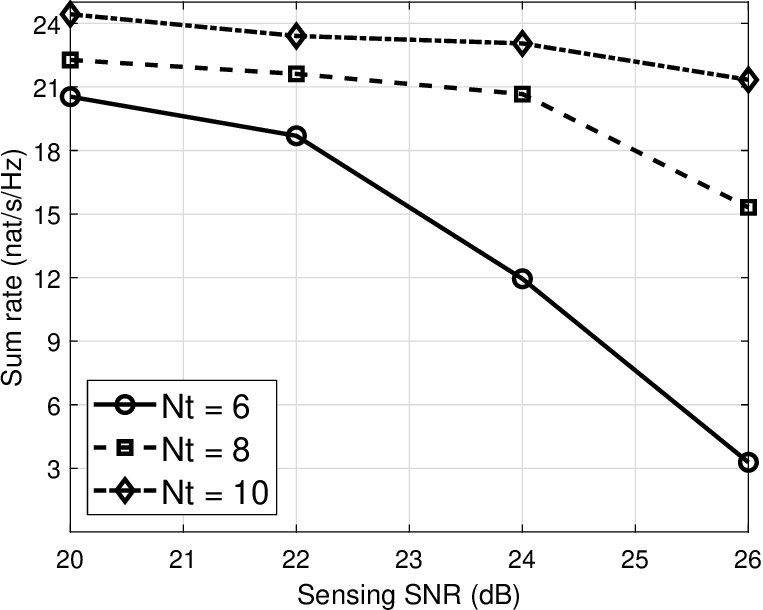}
	\caption{Sum rate versus different sensing SNR}
	\label{WSRvsSNR}
\end{figure}
Fig. \ref{WSRvsSNR} illustrates the relationship between sensing SNR and the sum rate, demonstrating how the sensing quality affects the overall system performance in a DP RIS-aided ISAC system. The simulation results are presented for three different values of $N_t$ ($N_t=6,8,10$). As the sensing SNR increases from 20 to 26 dB, the system experiences a degradation in sum rate performance across all configurations. Specifically, with $N_t=6$, the sum rate significantly decreases from 21 to 3 nat/s/Hz, showing the highest sensitivity to SNR changes. For $N_t=8$, the degradation is more moderate, declining from 22.5 to 15 nat/s/Hz, while $N_t=10$ demonstrates the most robust performance, with the sum rate decreasing from 24 to 21 nat/s/Hz. These results reveal a fundamental trade-off between sensing quality and communication performance in the ISAC system, where higher sensing SNR leads to reduced sum rates. The varying slopes of the curves suggest that larger $N_t$ values provide better resilience against this performance trade-off, offering important insights for system design optimization.

\subsection{Transmit beampattern}
\begin{figure}[t]
	\centering\vspace*{-0.50\baselineskip}
	\includegraphics[width=3.5in]{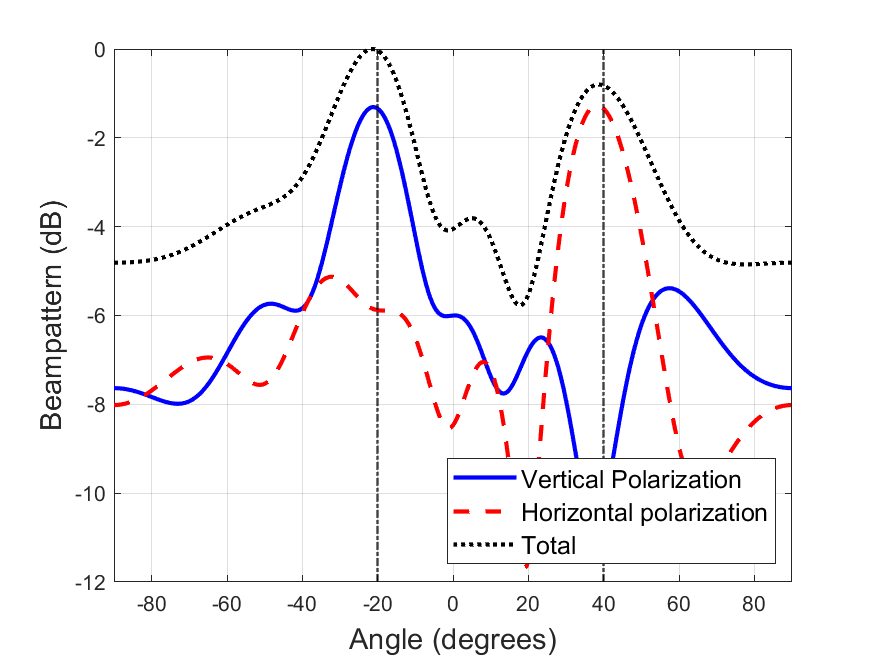}
	\caption{Transmit beampattern}
	\label{beam-ppattern}
\end{figure}
Fig. \ref{beam-ppattern} shows the synthesized far-field pattern with polarization. According to \cite{fuchs2011optimal}, the magnitude of pattern in vertical and horizontal polarization is calculated by 
\begin{align}
	&|P_v(\theta)|^2=|\mathbf{a}(\theta)^{\mathrm{H}}\mathbf{E}_v\mathbf{x}|^2,\\
	&|P_h(\theta)|^2=|\mathbf{a}(\theta)^{\mathrm{H}}\mathbf{E}_h\mathbf{x}|^2,\
\end{align}
where $\mathbf{E}_v$ and $\mathbf{E}_h$ is defined in \eqref{EvEh}, and $\mathbf{a}(\theta)$ is defined as
		\begin{equation}\label{aaa}
	\mathbf{a}(\theta) = \left[1, e^{j\frac{2\pi}{{\lambda}}d\sin \theta}, \ldots, e^{j\frac{2\pi}{{\lambda}}d(N_t-1)\sin \theta}\right]^{\mathrm{H}}.
\end{equation}
So the total beampattern of the array is
\begin{align}
	|P_{\text{total}}|^2=|P_v(\theta)|^2+|P_h(\theta)|^2.
\end{align}
From Fig. \ref{beam-ppattern}, it can be seen that different polarization beams are well directed to the targets, showing the good sensing performance of our algorithm. Different polarization beams point to different targets, which exploits the freedom of dual-polarization antennas for sensing compared to single-polarization arrays. 

\section{Conclusion}
In this paper, we considered a DP RIS-aided ISAC system under the DP channel. Our model features both the BS and RIS equipped with DP array elements, while users are equipped with a pair of DP antennas.
To address this problem, we first reformulated the sum rate maximization problem by using the WMMSE framework. We then applied an AO approach to iteratively optimize the phase-shift matrix $\mathbf{\Phi}$ and the beamforming matrix $\mathbf{W}$. Specifically, we developed an MM-based algorithm for optimizing $\mathbf{\Phi}$, and a penalty-based iteration method for $\mathbf{W}$.
Extensive simulation results validated the effectiveness of our proposed algorithm under various system configurations. 

		\bibliographystyle{IEEEtran}
		\bibliography{IEEEabrv,reference}

	\end{document}